\documentclass[journal]{IEEEtran}

\usepackage[caption=false,font=normalsize,labelfont=sf,textfont=sf]{subfig}

\usepackage{xcolor}
\usepackage{xspace}
\ifCLASSINFOpdf
\else
\fi
\usepackage[cmex10]{amsmath}

\usepackage{balance}

\usepackage{times}
\usepackage{fancyhdr,graphicx,amsmath,amssymb}
\usepackage[linesnumbered,ruled,vlined]{algorithm2e}
\let\oldnl\nl
\newcommand{\nonl}{\renewcommand{\nl}{\let\nl\oldnl}}
\usepackage{algorithmic}

\usepackage{fixltx2e}
\usepackage{dblfloatfix}
\usepackage{makecell}
\usepackage{booktabs}

\usepackage{cite}

\def\bz{{\bf z}}

\def\bO{{\bf O}}

\def\bW{{\bf W}}

\def\bh{{\bf h}}

\def\bQ{{\bf Q}}

\def\bw{{\bf w}}

\def\bI{{\bf I}}

\def\ba{{\bf a}}

\def\bs{{\bf s}}

\def\be{{\bf e}}

\usepackage{graphicx}
\usepackage{epstopdf}

\begin{document}




\title{Beamforming for Massive MIMO Aerial Communications: A Robust and Scalable DRL Approach}

 \author{\IEEEauthorblockN{Hesam Khoshkbari, Georges Kaddoum, Omid Abbasi, Bassant Selim and Halim Yanikomeroglu}
\thanks{H. Khoshkbari, G. Kaddoum and B. Selim are with the Department of Electrical Engineering, École de Technologie Supérieure, Montréal, Canada, e-mails:
\texttt{hesam.khoshkbari.1@ens.etsmtl.ca}, \texttt{Georges.Kaddoum@etsmtl.ca} and \texttt{bassant.selim@etsmtl.ca}. O. Abbasi and H. Yanikomeroglu are with the Department of Systems and Computer Engineering, Carleton University, Ottawa, Canada, e-mails: \texttt{omidabbasi@sce.carleton.ca} and \texttt{halim@sce.carleton.ca}. The corresponding author is H. Khoshkbari.

A preliminary version of this work was accepted for presentation at the 2025 IEEE International Conference on Communications (ICC) \cite{ICC}.}}

\IEEEaftertitletext{\vspace{0\baselineskip}}

\maketitle

{\color{black}\begin{abstract}
This paper presents a distributed beamforming framework for a constellation of airborne platform stations (APSs) in a massive Multiple-Input and Multiple-Output (MIMO) non-terrestrial network (NTN) that targets the downlink sum-rate maximization under imperfect local channel state information (CSI). We propose a novel entropy-based multi-agent deep reinforcement learning (DRL) approach where each non-terrestrial base station (NTBS) independently computes its beamforming vector using a Fourier Neural Operator (FNO) to capture long-range dependencies in the frequency domain. To ensure scalability and robustness, the proposed framework integrates transfer learning based on a conjugate prior mechanism and a low-rank decomposition (LRD) technique, thus enabling efficient support for large-scale user deployments and aerial layers. Our simulation results demonstrate the superiority of the proposed method over baseline schemes including WMMSE, ZF, MRT, CNN-based DRL, and the deep deterministic policy gradient (DDPG) method in terms of average sum rate, robustness to CSI imperfection, user mobility, and scalability across varying network sizes and user densities. Furthermore, we show that the proposed method achieves significant computational efficiency compared to CNN-based and WMMSE methods, while reducing communication overhead in comparison with shared-critic DRL approaches.
\end{abstract}
}
\IEEEpeerreviewmaketitle

\section{Introduction}  
\subsection{Overview}
Non-terrestrial base stations (NTBSs), which are characterized by ubiquitous connectivity, enhanced network reliability, sustainability, and supporting increasing data demand \cite{kurt2021vision,alam2021high,10355087,10417095} are widely regarded as essential components of future wireless communication systems. To support these evolving needs, the Third Generation Partnership Project (3GPP) has recently standardized non-terrestrial networks (NTN) in Releases 17 and 18 \cite{9904665,10355087}. Overall, airborne platform stations (APSs), including high-altitude platform stations (HAPS) and low-altitude platform stations (LAPS), offer several advantages. Compared  to satellites, they provide lower launch costs, reduced latency, quasi-stationary positioning, and lower path loss. Compared to unmanned aerial vehicles, they offer longer flight durations and wider coverage areas. However, to fully exploit the benefits of APSs, it is necessary to use a constellation of such platforms, which introduces new challenges in radio resource allocation, such as user association, beamforming, and interference management. Furthermore, any proposed resource allocation method should be sufficiently robust to the dynamic nature of NTN and scalable to support a constellation of NTBSs while minimizing signaling overhead and computational complexity.

To this end, in this paper, we propose a multi-agent deep reinforcement learning (DRL) beamforming method to maximize the network's sum rate using only local imperfect channel state information (CSI). The proposed approach incorporates entropy-based exploration \cite{haarnoja2018soft,haarnoja2018soft2} and employs the Fourier neural operator (FNO) \cite{li2020fourier,bonev2023spherical}, wherein beamforming is modeled as a continuous function of the CSI and FNO is used to learn this function in the frequency domain. Additionally, we incorporate transfer learning and low-rank decomposition to reduce computational complexity and improve the scalability of the proposed method.

\subsection{Related works}
In this section, we review related works on aerial communication, NTN, and the use of NTBSs in wireless communication.

Exploring power and sub-carrier allocation for interference management in a terrestrial network (TN) enhanced with a HAPS, the authors of \cite{shamsabadi2022handling} transformed the original problem into a convex one using a reformulation-linearization approach and proposed an iterative algorithm to solve it. The results revealed that HAPS integration improves spectral efficiency as compared to a TN-only deployment. In \cite{alsharoa2020improvement}, the authors formulated a joint optimization problem involving access and backhaul link association, along with power allocation in a three-layer VHetNet comprising terrestrial, aerial, and satellite components to maximize downlink throughput for ground users. User association in NTN was also investigated in various contexts \cite{10304250,9904576,10103832}. For instance, in \cite{10304250}, the authors suggested using HAPS as a backup BS to offload users from small base stations with low traffic, thereby reducing energy consumption. In \cite{9904576}, the authors employed a genetic algorithm to optimize user association for maximizing sum rate while minimizing mobility-induced handoffs. Meanwhile, in \cite{10103832}, aiming to minimize handover
occurrences, the authors examined the use of HAPS as a mobile computing center for vehicles transitioning between roadside units.
In another relevant study \cite{8892508}, the authors studied beamforming at the HAPS to manage the inter-layer interference between a HAPS and a satellite to improve network {\color{black}sum rate.}
Furthermore, joint optimization of user association and beamforming across TN and NTN layers was also addressed in several previous studies \cite{10304301,10445467,10379023}. Specifically, in \cite{10304301}, the authors considered a VHetNet composed of TN, HAPS, and a geostationary earth orbit (GEO) satellite, with the HAPS acting as a relay between the GEO satellite and ground users. In this context, users were first assigned to either terrestrial base stations (TBSs) or the GEO satellite (via the HAPS), and then beamforming at the HAPS and the power allocation at the single-antenna TBSs were determined using a weighted minimum mean square error (WMMSE) approach. Furthermore, in \cite{10445467}, the authors addressed the joint user association and beamforming problem to enhance spectral efficiency, applying an iterative method similar to \cite{shamsabadi2022handling} to solve the mixed-integer non-linear programming formulation. Another relevant study \cite{10379023} examined a network composed of TBSs and multiple hot-air balloons and performed 
user association and beamforming. The results highlighted the performance gains achieved with hot-air balloons over conventional TN-only architectures.

With the advent of artificial intelligence (AI), AI-based approaches have come to be extensively used to address resource allocation challenges in next-generation wireless communications. Particular attention was paid to DRL methods that are characterized by their adaptation capacity in dynamic wireless environments, independence from prior data and ability to operate without requiring global CSI. Consequently, following the demonstrated success of DRL for radio resource allocation in TNs \cite{alwarafy2021deep,feriani2021single}, several studies have investigated the application of DRL in NTN \cite{10016705,10599519}. Furthermore, in \cite{jo2022deep}, seeking to minimize the HAPS downlink outage probability, the authors addressed power allocation in a HAPS-integrated network by proposing a double deep Q-learning approach to mitigate inter-layer interference between a HAPS and a TN. A deep Q-learning (DQL)-based method was also introduced in \cite{cao2020deep} for user association in an NTN, aiming to maximize throughput while minimizing handoffs caused by NTBS mobility. In another relevant study \cite{10615903}, the authors examined the HAPS cell switching problem and proposed a Q-learning algorithm for energy-efficient cell switching decisions. {\color{black}Similarly, in \cite{10681113}, a graph neural network-based DRL framework was developed to capture spatial correlations in the wireless network and to make task offloading decisions to either HAPSs or satellites, depending on the delay tolerance associated with each task.} {\color{black}In \cite{10934003}, the authors investigated energy-efficient UAV-driven multi-access edge computing from a many-agent perspective. They proposed a distributed stochastic actor–critic algorithm with perturbed actors, modular networks, and reward shaping, demonstrating superior energy efficiency and safe UAV trajectory design compared to baseline MADRL methods.} {\color{black}In \cite{9748970}, the authors studied trajectory planning for cellular-connected UAVs and proposed a DRL framework enhanced with quantum-inspired experience replay (QiER). Using Grover-iteration–based amplitude amplification, QiER balances sampling priority and diversity, leading to faster convergence and more stable learning performance compared to conventional DRL and optimization baselines.} {\color{black}In \cite{10086561}, the authors addressed radio resource management in cellular-connected UAV networks, where UAVs coexist with terrestrial UEs under severe LoS-induced interference. They formulated a joint RB allocation and beamforming design problem and proposed a hybrid D3QN–TD3 solution, achieving effective outage minimization under realistic ITU building-based channel models.}
Last but not least, in our previous studies \cite{10171805,10330634,10312746,11059542}, we investigated user association strategies in MIMO HAPS-integrated networks to enhance the network's sum rate. Specifically, in \cite{10171805}, we developed a DQL approach for user association between a HAPS and a TBS, assuming that the agent operates solely with delayed CSI. This line of research was further extended in \cite{10330634} where we addressed the communication overhead caused by sharing CSI between users and the HAPS. In this work, we introduced a state-action-reward-state-action (SARSA)-based DRL method, where the agent only relies on delayed terrestrial CSI (i.e., between users and the TBS) for user scheduling. We also demonstrated that the SARSA framework yields better performance than DQL in scenarios with noisy CSI. Furthermore, in \cite{10312746}, we considered a more complex three-layer architecture consisting of a TBS, a HAPS, and a very low-earth orbit satellite, where the CSI available to the agent is limited to that of users’ previously associated BSs. This led to a 33$\%$ decrease in CSI sharing overhead and rendered the required CSI independent of the network layer within the VHetNet. The problem was modeled as a partially observable Markov decision process, and we proposed an action-specific deep recurrent Q-network approach to optimize the network's sum rate. Furthermore, we discussed how incorporating previous actions can improve learning performance in settings where global CSI is unavailable. {\color{black}In \cite{11059542}, we adopted convolutional neural network (CNN) layers to enhance the generalization capability of our DRL method, where each user acts as an agent and decides whether to connect with the TBS or HAPS. We showed that by using CNN layers and incorporating transfer learning, our model can be generalized over different numbers of users without fine-tuning.}

Along with NTN, several previous studies focused on studying deep neural network (DNN)-based beamforming in TNs. For instance, in \cite{9439874}, the authors introduced an unsupervised learning approach for hybrid beamforming in a single-cell setup. {\color{black}The proposed DNN simultaneously addressed classification by selecting analog beamforming vectors from a codebook, and regression by computing the digital beamforming components.} In another relevant study \cite{karkan2024sage}, a meta-learning-based DNN was proposed to achieve generalization across different antenna configurations along the x, y, and z axes of the base station. Meanwhile, in \cite{9112250}, the authors presented a hybrid beamforming framework for a single-cell scenario with one base station and one multi-antenna user. This framework employed a combination of deep deterministic policy gradient (DDPG) and WMMSE techniques to determine both analog and digital precoders for the base station and the user.

However, despite numerous studies on resource allocation in NTN and APS-integrated networks, distributed beamforming for a constellation of APSs under imperfect and local CSI remains an open problem. Another limitation of previous research is scarcity of studies on scalability of AI-based beamforming methods when faced with a large number of users and the increased computational complexity introduced by the growing number of BSs and layers in VHetNets. To address these challenges, we propose an entropy-based multi-agent DRL approach leveraging FNO, combined with transfer learning and low-rank decomposition, to maximize the network’s sum rate in a constellation of APSs. In what follows, the contributions of the present study are outlined in further detail.
{\color{black}\subsection{Motivation and contributions}
Beamforming in NTNs faces the following three critical challenges: 
(i) robustness against imperfect CSI arising from APS jittering, Doppler effects, and RF impairments \cite{10417095,khennoufa2025multi}; 
(ii) scalability with respect to the number of users, base stations, and aerial layers; and 
(iii) impractical dependency of traditional optimization-based solutions on global CSI \cite{10304301,10445467,10379023}, which leads to excessive computational cost and signaling overhead among APSs. Moreover, previous studies focus on the integration of a single aerial layer with terrestrial networks \cite{10304301,10445467,10379023,10171805,10330634,10312746}. Furthermore, as stated in \cite{shamsabadi2024interference}, conventional DRL methods in APS-integrated networks suffer from unstable training due to the stochastic and unpredictable nature of the environment, difficulty in handling continuous actions related to power allocation and beamforming, and poor exploration in large action spaces. To overcome these limitations, in this study, we propose a distributed stochastic beamforming strategy where each APS operates as an independent DRL agent that learns a robust and generalizable mapping from its own local imperfect CSI to continuous beamforming vectors. Our approach integrates entropy-driven exploration, Fourier neural operators (FNOs), conjugate-prior-based transfer learning, and low-rank decomposition (LRD) to achieve robustness, scalability, and computational efficiency across diverse NTN scenarios. {\color{black}Table \ref{tab:literature} summarizes how our work compares to relevant previous studies.}
\begin{table*}[t]
\renewcommand{\arraystretch}{1.5}
\caption{{\color{black}Comparison of the present study to previous research on APS-integrated (HAPS and LAPS)}}
\centering
\begin{tabular}{|c||c||c||c||c||c||c||c||c||c||c||c||c|}
\hline
  & \cite{10171805} & \cite{10330634} & \cite{10312746} & \cite{11059542} & \cite{10304301} & \cite{10445467} & \cite{shamsabadi2022handling} & \cite{10379023}  & \cite{alsharoa2020improvement} & \cite{10103832} & Present study \\
\hline
Multiple APSs &  &  & & & & & & \textbf{X} & \textbf{X}&  & \textbf{X} \\
\hline
Knowledge transfer between layers/APSs &  &  & & &  &  &  &  & & &\textbf{X} \\
\hline
Partial CSI &  & \textbf{X} & \textbf{X} & \textbf{X} & & & & &  & &\textbf{X} \\
\hline
Imperfect CSI-during training &  &  &  & & & & & &  & & \textbf{X} \\
\hline
Imperfect CSI-during test & \textbf{X} & \textbf{X} & \textbf{X} & & & & & &  & &\textbf{X} \\
\hline
Mobile users &  &  & & & & & & & & \textbf{X}&\textbf{X} \\
\hline
\end{tabular}
\label{tab:literature}
\end{table*}

The main contributions of this paper can be summarized as follows:
\begin{itemize}
\item \textbf{Distributed stochastic beamforming under local imperfect CSI:} We design a multi-agent entropy-based DRL framework where each APS (HAPS or LAPS) independently computes its beamforming vectors using only its local imperfect CSI during both training and test stages. In addition, our method accounts for mobile users, time-varying channels, Doppler shifts, and shadowing effects to address real-world deployment requirements.

\item \textbf{Operator learning via FNO:} The proposed DRL model integrates FNO, thus allowing each agent to learn a general mapping from CSI to beamforming weights in an infinite function space. Said differently, our DRL agents learn the underlying function, rather than specific numerical patterns, enabling them to capture long-range dependencies in the CSI and achieve robust generalization to unseen scenarios, regardless of fluctuations due to noise or mobility.

\item \textbf{Scalability through transfer learning and low-rank decomposition:} Our framework incorporates a transfer learning mechanism based on a conjugate-prior method to transfer the learned beamforming policy between aerial layers. This eliminates the need to retrain a separate DNN for each layer, which effectively reduces the computational burden introduced by the multi-layer structure of NTN. Furthermore, we integrate a low-rank decomposition technique into our method to effectively manage the large action space caused by the increasing number of users.

\item \textbf{Comprehensive robustness and scalability analysis:} We evaluate the framework under multiple practical conditions, including additive and multiplicative CSI noise models, the trade-off between using a replay buffer and regenerating data, different user velocities, and various numbers of clusters and users. The results reveal consistent superiority over WMMSE, zero-forcing (ZF), maximum ratio transmission (MRT), CNN-based DRL, and DDPG baselines in terms of sum rate, computational efficiency, and communication overhead, thereby validating robustness and scalability of our approach.  
\end{itemize}}
\textbf{Notations: }Upper-case boldfaced letters and lower-case boldfaced letters are used to represent matrices and vectors, respectively; notation    $(\cdot)^T$ denotes the transpose of a vector/matrix; ${\rm diag} (\bs) $ represents a diagonal matrix with diagonal entries being the elements of vector  $\bs$. Calligraphic fonts (e.g., ${\cal A}$) are used to signify sets, where the cardinality of a set  ${\cal A}$ is denoted as $|\cal A|$. $\mathbb{E}\{ \cdot \} $ denotes the mathematical expectation; 
 $[.]_u$  extracts the $u$-th row of a matrix. Finally,  $|| \cdot||$ is used for the Frobenius norm.

\section{system model and problem formulation}\label{sec:system model}
\begin{figure}[h]
	\centering
	\includegraphics[scale=0.1]{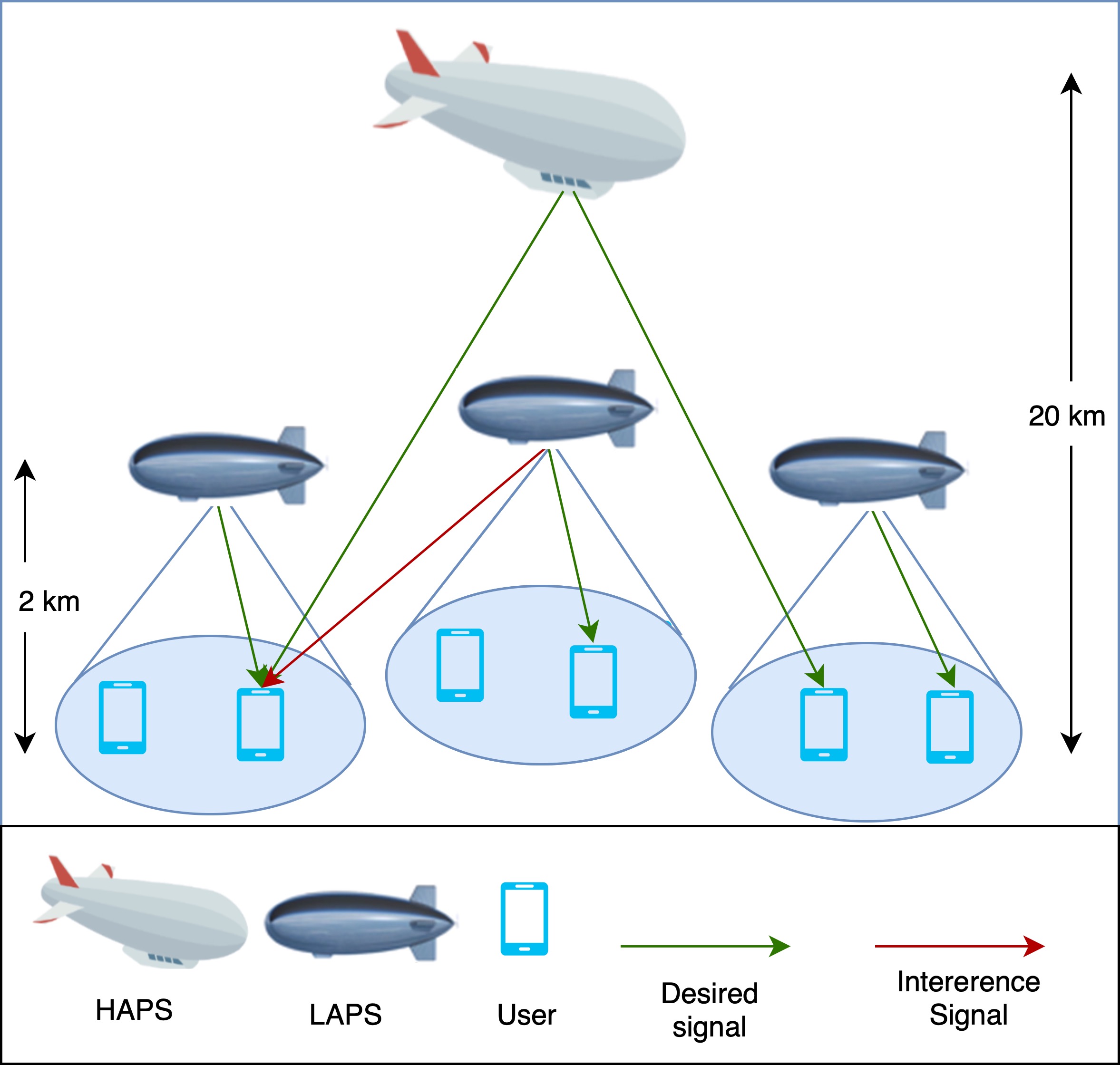}
	\caption{System model illustration for $B = 3$ LAPSs and $K = 2$ users per cluster. }
	\label{fig:systemmodel}
\end{figure}

In this study, we propose a distributed beamforming approach designed for a constellation of APSs, with the objective of maximizing the downlink sum rate in the network. We consider a two-tier NTN massive MIMO architecture where the set of APSs is denoted as $\mathcal{B} = \{b_{0}, b_{1},b_{2},\ldots,b_{B} \}$, with $b_0$ representing the HAPS. Each APS is equipped with $N_b$ antennas. The lower tier consists of $B$ LAPSs, where each LAPS serves the users located within its respective cluster over a shared frequency band. A collection of single-antenna users is defined as $\mathcal{U} = \{u_{1},u_{2},\ldots,u_{U} \}$, with $U = K \times B$, where $K$ users are uniformly distributed per cluster and where $N_b \geq K$. The upper tier comprises a HAPS equipped with $N_{b_0}$ antennas and operating on a frequency band distinct from that of the LAPSs. {\color{black} We assume each APS employs a uniform planar array (UPA) with half-wavelength spacing in both horizontal and vertical dimensions.} To exploit the benefits of line-of-sight (LoS) links, we adopt a dual connectivity model, wherein each user can simultaneously receive service from both the HAPS and its associated LAPS. This design is supported by recent evidence showing that, in the presence of LoS channels, the network's sum rate increases with multi-connectivity \cite{10643015}. Moreover, as shown in \cite{10530195}, multi-connectivity enhances coverage probability and average data rates in NTN. {\color{black}In our framework, the two layers do not explicitly coordinate their scheduling decisions; instead, each APS operates as an independent agent with access to its local CSI only, while implicit coordination emerges through the shared system-level reward during training.} The system configuration for $B = 3$ LAPSs and $K = 2$ users per cluster is shown in Fig. \ref{fig:systemmodel}

\subsection{Channel model}
At each time slot $t$, the wireless channel between user $u$ and BS $b$ is expressed as follows:

\begin{equation}
\bh^{t}_{b,u} = \widehat{\bh}_{b,u}^{t} \sqrt{L_{b,u}^{t}},  
    \label{eq:TBS channel}
\end{equation}
where $\bh_{b,u}^{t}$ denotes the $1 \times N_{b}$ channel vector from user $u$ to the antennas of BS $b$. Here, $\widehat{\bh}_{b,u}^{t}$ and $L_{b,u}^{t}$ represent the small-scale and large-scale fading components, respectively. {\color{black}The large-scale fading is given by:}

\begin{equation} \label{eq:largescaleTBS}
    \log_{10} L_{b,u}^{t} = \log_{10} \left(\frac{c}{4 \pi f_{c_{b}} d_{b,u}^{t}}\right)^2 - \psi_{d B},
\end{equation} 
where $f_{c_{b}}$ is the carrier frequency for BS $b$, $c$ is the speed of light, $d_{b,u}^{t}$ is the distance between user $u$ and BS $b$ at time $t$, and, finally, $\psi_{d B}$ is a zero-mean Gaussian random variable with variance $\sigma_{\psi_{d B}}^2$ representing shadowing effects.

Due to the LoS nature of APS links, the small-scale fading component $\widehat{\bh}_{b,u}^{t}$ follows a Rician model, expressed as:

\begin{equation}
  \widehat{\bh}_{{b,u}}^{t} = \sqrt{\frac{X}{1+X}} \widehat{\bh}_{{b,u}_{LOS}}^{t}+\sqrt{\frac{1}{1+X}} \widehat{\bh}_{{b,u}_{NLOS}}^{t}, \label{eq:LosNLos}
\end{equation}
where $\widehat{\bh}_{{b,u}_{LOS}}^{t}$ and $\widehat{\bh}_{{b,u}_{NLOS}}^{t}$ denote the LoS and non-LoS (NLoS) components, respectively, while $X$ is the Rician factor.

To incorporate Doppler effects, we adopt Jakes' model \cite{dent1993jakes} for the NLoS fading component as follows:

\begin{equation}
\widehat{\bh}_{{b,u}_{NLOS}}^{t} \triangleq   \rho \widehat{\bh}_{{b,u}_{NLOS}}^{t-1} + \sqrt{ 1 -{\rho ^2}} \bz_{{b,u}}^{t}, \quad for \quad u = 1 , ..., U, 
\label{eq:HAPS NLOS}
\end{equation}
where $\bz_{{b,u}}^{t}$ is a complex Gaussian noise vector and $\rho$ is the correlation coefficient determined by the Doppler frequency \cite{abou2005binary}. As discussed in \cite{falletti2006integrated}, Jakes’ model is appropriate to simulate fading in APS communication channels.

{\color{black}The LoS component $\widehat{\bh}_{{b,u}_{LOS}}^{t}$ is modeled as follows:}
\begin{equation}
\widehat{\bh}_{{b,u}_{LOS}}^{t} = \mathbf{a}\left(\theta_{b,u}, \phi_{b,u}\right) \otimes \mathbf{b}\left(\theta_{b,u}, \phi_{b,u}\right), \label{eq:HAPS LOS}
\end{equation}
where $\otimes$ denotes the Kronecker product. Vectors $\mathbf{a}$ and $\mathbf{b}$ capture the phase progression across antenna elements in the horizontal and vertical dimensions, respectively, and are given by:
\begin{equation}
\begin{split}
    \mathbf{a}\left(\theta_{b,u}, \phi_{b,u}\right)=\left[1, e^{j 2 \pi d_h}, \ldots, e^{j 2 \pi\left(\sqrt{N_{b}}-1\right) d_h}\right]^T, \\
    \mathbf{b}\left(\theta_{b,u}, \phi_{b,u}\right)=\left[1, e^{j 2 \pi d_v}, \ldots, e^{j 2 \pi\left(\sqrt{N_{b}}-1\right) d_v}\right]^T,
\end{split}
\end{equation}
where $\theta_{b,u}$ and $\phi_{b,u}$ denote elevation and azimuth angles of user $u$ relative to BS $b$. Defining the wavelength as $\lambda = c / f_{c}$, the horizontal and vertical phase shifts are computed as $d_h = d_{\mathrm{x}} \cos \theta_{b,u} \sin \phi_{b,u} / \lambda$ and $d_v = d_{\mathrm{y}} \cos \theta_{b,u} \cos \phi_{b,u} / \lambda$, assuming antenna element spacing $d_{\mathrm{x}} = \lambda / 2$ and $d_{\mathrm{y}} = \lambda / 2$, respectively.

{\color{black}\subsection{User mobility model}
In this study, as outlined in Section \ref{sec:method}, we adopt an episodic training scheme where the simulation environment resets at the beginning of each episode spanning $T$ time slots. At the start of every episode, each user uniformly and randomly selects a movement direction $\Theta_{u} \in [0, 2\pi]$ and maintains a constant velocity $v$ throughout the episode. Based on the user’s current position $(x_{u}^{t}, y_{u}^{t})$ at time slot $t$, the location at time $t+1$ is computed as follows: 

\begin{equation} 
\begin{split} 
x_{u}^{t+1} = x_{u}^{t} + D_{\max} \cos \Theta_{u}, \\ 
y_{u}^{t+1} = y_{u}^{t} + D_{\max} \sin \Theta_{u}, 
\end{split} 
\end{equation} 
where $D_{\max} = v T_{c}$ represents the maximum displacement during a time slot, while $T_{c}$ is the duration of each slot. We assume that, throughout the episode, users remain within their respective clusters. If a user encounters a cluster boundary, it randomly selects a new direction and continues to move while staying confined to the cluster area. As users move, $d_{b,u}^{t}$ changes with time leading to the time variant large scale fading in Eq. \eqref{eq:largescaleTBS}. This Random Direction Mobility Model with constant velocity is consistent with 3GPP evaluation scenarios and was widely adopted in previous studies \cite{9904576,10436904,10103832,9931560,10680056}.}

\subsection{Rate expression}
We assume that each layer in the network operates over a separate frequency band, thereby eliminating the risk of inter-layer interference. The downlink transmission rate from BS $b$ to user $u$ at time slot $t$ is defined as follows:
\begin{equation} \label{eq:rate} 
R_{u}^{t} = \log_{2} (1 + \text{SINR}_{b,u}^{t}), 
\end{equation}
where $\mathrm{SINR}_{b,u}^{t}$ represents the downlink signal-to-interference-plus-noise ratio (SINR) for the link between BS $b$ and user $u$. For the case where $b \neq b_0$, the SINR is computed as follows:
\begin{equation}
    \text{SINR}_{b,u}^{t} = \frac{| \bh^{t}_{b,u} \bw_{b,u}^{t}  |^2}{\sum_{\substack{b' \in \mathcal{B} \\b^{\prime} \neq b_0 }} \sum_{\substack{u^{\prime} \\u^{\prime} \neq u} } | \bh^{t}_{b',u} \bw_{b',u'}^{t} |^2+\sigma^{2}}, \label{eq:SINR_TBS}
\end{equation} 
where $\bw_{b,u}^{t}$ is the $N_b \times 1$ beamforming vector transmitted from LAPS $b$ to user $u$ at time $t$. In contrast, when $b = b_0$ (i.e., the HAPS), the SINR is obtained as:
\begin{equation}
    \text{SINR}_{b,u}^{t} = \frac{| \bh^{t}_{b,u} \bw_{b,u}^{t}  |^2}{\sum_{\substack{u^{\prime} \\u^{\prime} \neq u} } | \bh^{t}_{b,u} \bw_{b,u'}^{t} |^2+\sigma^{2}}. \label{eq:SINR_HAPS}
\end{equation} 

\subsection{Problem formulation}
Let $\mathcal{U}_{b}$ denote the set of users in cluster $b$ served by the corresponding LAPS. For the sake of notational simplicity, we omit the time index and define the optimization problem as follows: \begin{subequations}\label{eq:opt}
\begin{alignat}{4}
\max_{{\bw_{b,u}}} & \quad \sum_{u \in \mathcal{U}} R_{u} \tag{\ref{eq:opt}}\\
\text{subject to} & \quad \sum_{u \in \mathcal{U}_{b}} \left\|\mathbf{w}_{b,u}\right\|_2^2 \leq P_{\max}^{b}, \hspace{0.2cm} \forall b, b \neq b_0 \label{eq:opta}\\
&  \quad \sum_{u \in \mathcal{U}} \left\|\mathbf{w}_{b,u}\right\|_2^2 \leq P_{\max}^{b}, \hspace{0.2cm} b=b_0 \label{eq:optb} \\ 
&  \quad\quad \cap_{b=1}^{B} \mathcal{U}_{b} = \emptyset \label{eq:optc}\\
& \quad \bw_{b,u} \in \mathbb{C}^{N_b \times 1}, \hspace{0.2cm}  \forall b \label{eq:optd}.
\end{alignat}
\end{subequations}

In this formulation, $P_{\max}^{b}$ denotes the power budget at BS $b$. The first two constraints, \eqref{eq:opta} and \eqref{eq:optb}, ensure that the transmission power used by each BS remains within its maximum supply. Constraint \eqref{eq:optc} enforces non-overlapping user association, where each user is served exclusively by the LAPS within its cluster. Since the full and global CSI is unavailable in our setup, we tackle the problem in \eqref{eq:opt} using a multi-agent entropy-driven DRL approach, where each BS (LAPS or HAPS) acts as an autonomous agent responsible for computing its beamforming vectors.
{\color{black}\section{Proposed distributed beamforming algorithm}\label{sec:method} 

Solving the optimization problem in Eq. \eqref{eq:opt} through optimization techniques \cite{10304301,10445467,10379023} and mathematical-based methods, such as WMMSE, ZF, or MRT, require global CSI and incurs prohibitive computational cost, which rapidly scales with the number of users, antennas, and APS layers. Therefore, these methods are unsuitable for large NTN deployments where CSI is imperfect, user mobility induces rapid channel variations, and distributed scalability is essential. To overcome these limitations, in this study, we adopt a multi-agent DRL framework that enables each APS to operate as an autonomous agent computing its own beamforming vectors from local CSI without relying on centralized coordination.  

In contrast to classical DRL methods based on the $\varepsilon$-greedy strategy that often struggle with continuous actions, and large action spaces \cite{jo2022deep,cao2020deep,10615903,10171805,10330634,10312746}, we design a stochastic actor-only policy parameterized by Gaussian distributions. This policy uses entropy-based exploration \cite{haarnoja2018soft,haarnoja2018soft2} to encourage diverse yet structured action sampling. Moreover, such entropy-based stochastic DRL also demonstrated improved robustness to hyper-parameter tuning as compared to DDPG \cite{duan2016benchmarking,henderson2018deep,10247079}. Furthermore, instead of relying on CNNs or RNNs that primarily capture local features \cite{10415840,10695364,bonev2023spherical}, we use FNOs to learn a functional mapping from CSI to beamforming weights in the frequency domain, enabling the capture of global dependencies and improving generalization under noisy or time-varying conditions. Finally, to ensure scalability, our framework incorporates a conjugate-prior-based transfer learning mechanism that reuses policies across aerial layers without retraining, as well as a low-rank decomposition strategy reducing output dimensionality in massive MIMO settings.}
\subsection{Fourier neural operator} \label{sec:fno}
To learn a robust and generalizable mapping from CSI to beamforming weights in NTN environments, we use the FNO, a mesh-independent operator learning framework that maps between infinite-dimensional function spaces \cite{li2020fourier,gopakumar2023fourier}. Unlike traditional CNNs that learn mappings from finite-dimensional grids and largely rely on local spatial correlations, FNOs use frequency-domain representations that enable them to capture global patterns and long-range dependencies across the CSI domain \cite{bonev2023spherical,liu2025enhancing}.
An FNO layer applies the following transformation: 
\begin{equation} 
y = \sigma\left(\mathcal{F}^{-1}\left(Z \cdot \mathcal{F}(x)\right) \right), 
\end{equation} 
where $x$ is the input (e.g., real and imaginary CSI components), $\mathcal{F}$ and $\mathcal{F}^{-1}$ denote the Fast Fourier Transform (FFT) and its inverse (IFFT), $Z$ is a learnable filter applied to the selected frequency modes, and finally, $\sigma$ is a non-linear activation function (e.g., Rectified Linear Unit (ReLU)).

In each FNO layer, the input $x$ is first transformed to the frequency domain using $\mathcal{F}(x)$. Then, a fixed number of dominant frequency modes ($n_{\text{modes}_x}$ and $n_{\text{modes}_y}$ in a 2D signal) is retained, while the remaining frequency modes are truncated, reducing noise and irrelevant fluctuations. Learnable tensor $Z$ operates on these modes, capturing meaningful spectral correlations. This filter is applied with $\overline{L}$ layers in $\overline{O}$ channels. The filtered representation is then mapped back to the time domain using $\mathcal{F}^{-1}$.

This structure allows FNO to learn a continuous operator $f:\text{CSI} \mapsto \text{beamforming}$, generalizing across varying user positions, Doppler shifts, and imperfect CSI. Its frequency-based abstraction also mitigates instability caused by user mobility and noise, thus making FNO particularly suitable for real-world NTNs.

\subsection{Action} \label{sec:action}
The action space corresponding to BS $b$ is defined as follows:
\begin{equation} \label{eq:action}
   \mathcal{A}_{b}= \begin{cases} \{ \Re(\bw_{b,u}), \Im(\bw_{b,u}) \} \hspace{2mm} \text {for} \hspace{2mm} u \in \mathcal{U}_{b}, \forall b, b \neq b_0
    \\   \{ \Re(\bw_{b,u}), \Im(\bw_{b,u}) \} \hspace{2mm} \text {for} \hspace{2mm} u \in \mathcal{U}, b = b_0,\end{cases} 
\end{equation}
where $\Re(\bw_{b,u})$ and $\Im(\bw_{b,u})$ denote the real and imaginary components of the complex beamforming vector $\bw_{b,u}$, respectively.
Since $\bw_{b,u}$ is inherently a complex-valued vector and neural networks are typically designed to process real-valued data, its real and imaginary parts should be handled independently.
\subsection{State}
For each BS, at time slot $t$, the state is defined as follows:
\begin{equation} \label{eq:state}
   \bs_{b}^{t}= \begin{cases} [\tilde{\bh}_{b,u}^{t}] \hspace{2mm} \text {for} \hspace{2mm} u \in \mathcal{U}_{b}, \forall b, b \neq b_0,
    \\   [\tilde{\bh}_{b,u}^{t}] \hspace{2mm} \text {for} \hspace{2mm} u \in \mathcal{U}, b = b_0,\end{cases} 
\end{equation} 
where $\tilde{\bh}_{b,u}^{t}$ represents the imperfect CSI.

In this study, we examine the robustness of our proposed method under multiple models of CSI imperfection. To this end, we incorporate both additive and multiplicative noise, as well as Gaussian and non-Gaussian noise distributions.
Under the additive noise model, the imperfect CSI is expressed as follows:
\begin{equation} \label{eq:additive}
    \tilde{\bh}_{b,u}^{t}=\xi \bh_{b,u}^{t}+\sqrt{1-\xi^2} \be,
\end{equation}
where $\be$ is a complex Gaussian error vector and $\xi$ reflects the reliability of the channel estimation.
For the multiplicative noise case, the imperfect CSI is modeled using the following formulation:
\begin{equation} \label{eq:multiplicative}
     \tilde{\bh}_{b,u}^{t} = \bh_{b,u}^{t} \times \be, \hspace{2mm} \be \sim \Gamma(k', \theta'),
\end{equation} 
where $\Gamma(k', \theta')$ denotes the Gamma noise distribution, with $k'$ and $\theta'$ shape and scale parameters, respectively.

\subsection{Reward}
The agents are trained using a shared reward signal, which is defined as follows:
\begin{equation}
    r^{t} = \frac{1}{U}\sum_{u \in \mathcal{U}} R_{u}^{t}, \label{eq:reward}
\end{equation}
where $R_{u}^{t}$ represents the downlink rate for user $u$ as described in \eqref{eq:rate}, while $U$ denotes the total number of users in the network.

\subsection{Proposed entropy-based multi-agent DRL approach}
In this study, we adopt a centralized training and distributed execution framework, where two separate actor DNNs are trained, i.e. one for the LAPSs and another for the HAPS. The beamforming policy for each LAPS is represented by $\pi_{\boldsymbol{\Omega}}\left(\ba_{b}^{t} \mid \bs_{b}^{t}\right)$, while the HAPS beamforming policy is given by $\overline{\pi}_{\overline{\boldsymbol{\Omega}}}\left(\ba_{b_0}^{t} \mid \bs_{b_0}^{t}\right)$. Here, $\boldsymbol{\Omega}$ and $\overline{\boldsymbol{\Omega}}$ denote trainable parameters for the LAPS and HAPS actor networks, respectively. {\color{black}During execution, each BS inputs its current state \eqref{eq:state} into its associated actor to compute the corresponding beamforming vector.}

\begin{figure}[h]
	\centering
	\includegraphics[scale=0.088]{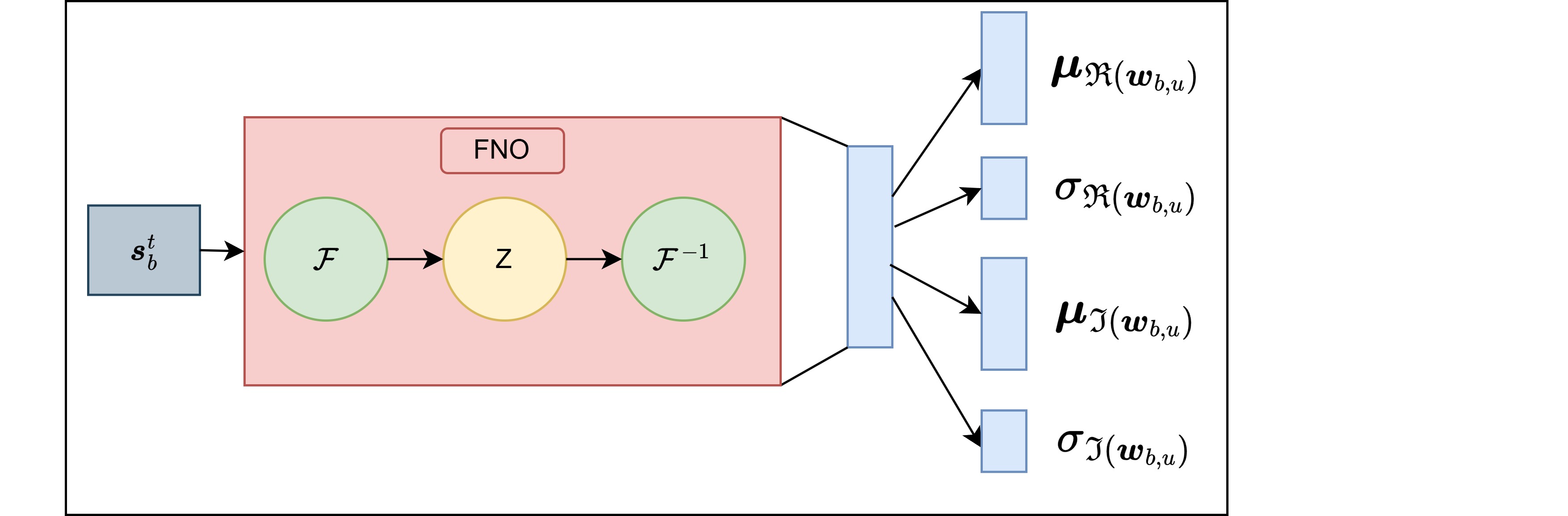}
	\caption{Architecture of the actor networks used for beamforming policy learning. Both the LAPS and HAPS actor networks share the same structure, consisting of one FNO layer to extract global dependencies from the CSI, followed by one hidden layer and four output heads corresponding to the means and variances of the real and imaginary parts of the beamforming vector. The beamforming action is sampled from a Gaussian distribution parameterized by these outputs. The only difference between the LAPS and HAPS networks lies in their output dimensions.}
	\label{fig:TBS actor}
\end{figure}

The architecture of the actor networks is shown in Fig.~\ref{fig:TBS actor}. Both networks follow the same structure, differing only in their output dimensions. To separately encode the real and imaginary  parts of the CSI, the input state, defined in \eqref{eq:state}, is formatted as a matrix with two channels. The output of each actor includes the following four components: $\boldsymbol{\mu}_{\Re(\boldsymbol{w}_{b,u})}$, ${\sigma}_{\Re(\boldsymbol{w}_{b,u})}$, $\boldsymbol{\mu}_{\Im(\boldsymbol{w}_{b,u})}$, and ${\sigma}_{\Im(\boldsymbol{w}_{b,u})}$, corresponding to the mean and standard deviation of the real and imaginary parts, respectively.
The beamforming vector is then sampled from the learned Gaussian distributions: $\Re(\bw_{b,u}) \sim \mathcal{N}(\boldsymbol{\mu}_{\Re(\boldsymbol{w}_{b,u})}, e^{{\sigma}_{\Re(\boldsymbol{w}_{b,u})}}.\bI)$ and $\Im(\bw_{b,u}) \sim \mathcal{N}(\boldsymbol{\mu}_{\Im(\boldsymbol{w}_{b,u})}, e^{{\sigma}_{\Im(\boldsymbol{w}_{b,u})}}.\bI)$. Consequently, the beamforming policies $\pi_\Omega\left(\ba_{b}^{t} \mid \bs_{b}^{t}\right)$ and $\overline{\pi}_{\overline{\boldsymbol{\Omega}}}\left(\ba_{b_0}^{t} \mid \bs_{b_0}^{t}\right)$ can be represented as $f_{\Omega}(\boldsymbol{\epsilon}^{t};\bs_{b}^{t})$ and $\overline{f}{\overline{\Omega}}(\boldsymbol{\overline{\epsilon}}^{t};\bs_{b_0}^{t})$, where $\boldsymbol{\epsilon}^{t}$ and $\boldsymbol{\overline{\epsilon}}^{t}$ are sampled from Gaussian distributions defined by the learned means and variances of the real and imaginary components.

The loss functions for the actor networks are defined as follows:
\begin{equation} \label{eq:actor loss}
\begin{split}
J_{\pi}(\boldsymbol{\Omega}) = \mathbb{E}_{(\bs_{b}^{t}, \ba_{b}^{t}, r^{t} ) \sim \mathcal{D}} \left[ \gamma \log(f_{\Omega}(\boldsymbol{\epsilon}^{t};\bs_{b}^{t})) - r^{t}   \right], \\
J_{\overline{\pi}}(\overline{\boldsymbol{\Omega}}) = \mathbb{E}_{(\bs_{b_0}^{t}, \ba_{b_0}^{t}, r^{t} ) \sim \mathcal{D}} \left[ \gamma' \log(\overline{f}_{\overline{\Omega}}(\boldsymbol{\overline{\epsilon}}^{t};\bs_{b_0}^{t})) - r^{t}   \right],
\end{split}
\end{equation} 
where $\mathcal{D}$ represents the replay buffer storing past transitions to reduce sample correlation and enhance training stability \cite{mnih2015human}. However, as discussed in Section \ref{sec:results}, our method learns the functional relationship between CSI and beamforming, enabling actor networks to update their parameters using synthetically generated CSI rather than relying solely on stored data. This solution is particularly advantageous when the network operator prefers not to share real CSI during training.
In these formulations, $\gamma$ and $\gamma'$ are hyperparameters governing the balance between exploration and exploitation. In the next section, we describe our proposed transfer learning approach, which is integrated into the DRL framework to eliminate the need to retrain a separate DNN for each layer for the HAPS. This mechanism effectively reduces computational complexity caused by the multi-layer structure of aerial communications.

\subsection{Proposed incorporated transfer learning} \label{sec:transfer}
While training individual DNNs for each layer in aerial communications may be justified in scenarios with fundamentally distinct tasks, this approach proves to be inefficient when both layers share the same architecture and objective—namely, beamforming. The DNN for the LAPSs handles both intra- and inter-layer interference, making its task inherently more complex than that of the HAPS, which operates under simpler conditions. In order to reduce the computational overhead, we propose a lightweight transfer learning technique where only the LAPSs’ DNN, referred to as the source network, is trained, and its learned parameters are transferred to the HAPS’ DNN, the target network, without requiring backpropagation through each layer.

Both networks share an identical structure: one FNO layer, a hidden layer, and two output layers. Since the FNO component operates in the frequency domain and captures global dependencies, it should be trained independently for each layer. However, the hidden and output layers are amenable to parameter transfer. We treat the trained weights of the LAPSs’ DNN as a prior distribution and the initial weights of the HAPS’ DNN as the likelihood. {\color{black}Assuming Gaussian priors and Gaussian likelihoods, which form a conjugate prior formulation, the resulting posterior is also Gaussian \cite{raiffa2000applied}.}
Let $\mathcal{L} = \{l | \hspace{1mm} \text{for} \hspace{1mm} l = \text{hidden layer}, \boldsymbol{\mu}_{\Re(\boldsymbol{w}_{b,u})}, {\sigma}_{\Re(\boldsymbol{w}_{b,u})}, \boldsymbol{\mu}_{\Im(\boldsymbol{w}_{b,u})}, {\sigma}_{\Im(\boldsymbol{w}_{b,u})} \}$ denote the set of transferable layers. The means and variances of the $l$-th layer for the LAPS and HAPS DNNs are denoted by $\mu_{\text{LAPS}_l}$, $\sigma_{\text{LAPS}_l}^2$, $\mu_{\text{HAPS}_l}$ and $\sigma_{\text{HAPS}_l}^2$, respectively. Assuming Gaussian distributions, the posterior distribution for the $l$-th layer of the HAPS DNN remains Gaussian, with mean $\mu_{\text{p}_{\text{HAPS}_l}}$ and variance $\sigma_{\text{p}_{\text{HAPS}_l}}^2$ given by:
\begin{equation} \label{eq:conjtrf}
\begin{split}
\sigma_{\text{p}_{\text{HAPS}_l}}^2=\left(\frac{1}{\beta \sigma_{\text{LAPS}_l}^2}+\frac{1}{\sigma_{\text{HAPS}_l}^2}\right)^{-1}, \\ \mu_{\text{p}_{\text{HAPS}_l}}=\sigma_{\text{p}_{\text{HAPS}_l}}^2\left(\frac{\beta \mu_{\text{LAPS}_l}}{\sigma_{\text{LAPS}_l}^2}+\frac{\mu_{\text{HAPS}_l}}{\sigma_{\text{HAPS}_l}^2}\right). 
\end{split}
\end{equation}

Here, $\beta > 0$ is a hyperparameter that regulates the influence of the LAPS prior on the posterior, and it is learned directly from the LAPS DNN. {\color{black}This is because, during the training phase, the LAPS agents may need to perform more extensive exploration in order to manage interference, a behavior that may not be required at the HAPS level.} This Bayesian-inspired update allows for an effective knowledge transfer across aerial layers while maintaining both robustness and scalability in distributed multi-agent systems.

\subsection{Proposed incorporated low-rank decomposition} \label{sec:lrd}
In this section, we present the low-rank decomposition (LRD) strategy adopted to support scalable beamforming design at the HAPS node. While the focus is on HAPS for clarity, the same decomposition technique is also applicable to the LAPSs.

Let $\bW \in \mathbb{C}^{U \times N_{b0}}$ represent the beamforming matrix at the HAPS, where $U$ denotes the total number of users and $N_{b0}$ is the number of antennas. In massive MIMO systems, as $U$ increases, the size of the beamforming matrix—and, consequently, the dimensionality of the DNN's output layer—increases proportionally. This rise in dimensionality leads to a higher computational cost, requires larger hidden layers, and increases the risk of overfitting. To effectively address these challenges, we approximate the full matrix $\bW$ as the product of the following two lower-dimensional matrices: 
\begin{equation} \label{eqlrd}
    \bW = \bQ \times \bO,
\end{equation} 
where $\bQ \in \mathbb{C}^{U \times j}$ and $\bO \in \mathbb{C}^{j \times N_{b0}}$ denote user-specific and antenna-specific beamforming components, respectively. In this context, the hyperparameter $j \ll \min(U, N_{b0})$ defines the dimension of the latent subspace used for this decomposition.

This factorized representation reduces the output size of the DNN from $U \times N_{b0}$ to $(U \times j) + (j \times N_{b0})$ for both real and imaginary parts. Consequently, the training and inference complexity is significantly decreased. In addition, by constraining the beamforming matrix $\bW$ to a low-rank manifold, this structure acts as a form of implicit regularization, with an effective rank not exceeding $j$.
Although applying a low-rank constraint may introduce suboptimality in cases where the optimal $\bW$ has a high rank, it remains a valid and practical approximation in many real-world NTN scenarios. In particular, APS-to-ground channels are typically dominated by LoS propagation, which results in a low-rank channel structure. Most of the signal energy is concentrated in the top singular modes. Therefore, retaining only the top $j$ modes through LRD preserves most of the beamforming performance while enhancing robustness to noise and improving system scalability.

\subsection{Training procedure} \label{sec:training}
Algorithm \ref{alg:TrainSAC} outlines the training procedure for our proposed framework. To capture long-range dependencies between users and antennas and to learn the operator mapping from CSI to beamforming, the input state is first reshaped into a two-channel matrix and then processed through a single FNO layer (see Fig.~\ref{fig:TBS actor}). The extracted features are subsequently flattened and fed into a fully connected layer, followed by the output layers to generate the action.

Both FNO layers for the HAPS and LAPS actors are configured with 8 output channels. For the two-dimensional Fourier transform (2D FFT), we retain 4 and 12 frequency modes along the height and width dimensions for the LAPS, and 8 and 20 modes for the HAPS. The fully connected layer contains a total of 512 neurons. All network weights are initialized using a Gaussian distribution. The replay buffer $\mathcal{D}$ is also initialized. ReLU activation functions are employed across all layers, except for the output layers where a softmax activation is applied.

To prevent overfitting to specific network conditions, we randomize the environment at the beginning of each episode by sampling initial channel states from a Gaussian distribution, randomly distributing users within their respective clusters and assigning each user a random movement direction. Each BS receives its corresponding state matrix and outputs the real and imaginary parts of the beamforming vector, either directly or in the latent subspace as described in Section \ref{sec:lrd}. The generated beamforming vectors are then normalized to satisfy the power constraints in \eqref{eq:opta} and \eqref{eq:optb}.

After observing the reward, the training proceeds in one of the following two ways: (i) the state matrices are stored in the replay buffer $\mathcal{D}$ and, at every $\eta$ time slot, a mini-batch is sampled for updating the actor networks, or (ii) at every $\eta$ time slot, a new mini-batch of synthetic channels is generated with randomly repositioned users to update the actor networks. As detailed in Section \ref{sec:results}, the functional nature of the learned mapping allows the network to update from unseen data, thus giving network operator the flexibility to choose between replay buffer or data regeneration.
In scenarios where transfer learning is applied, only the FNO layer of the HAPS DNN is trained, while the remaining layers are updated following the approach described in Section \ref{sec:transfer}. Finally, the fully trained actor networks are saved for evaluation during the testing phase.
\begin{algorithm}  
  \nonl \textbf{Input}:  Replay buffer $\mathcal{D}$, LAPS actor network parameters $\boldsymbol{\Omega}$, HAPS actor network parameters $\overline{\boldsymbol{\Omega}}$, system model defined in Section \ref{sec:system model}\;  
  Initialize buffer $D$ and weights $\boldsymbol{\Omega}$, $\overline{\boldsymbol{\Omega}}$.\\
  \For{ $i \in 1 \hspace{2mm} \text{to} \hspace{2mm} I_{\rm{episode}}$}{
      Distribute users, generate the initial channel values, and select the movement direction. \\
      \For{$t \in 1 \hspace{2mm} \text{to}  \hspace{2mm} T$}{
           
        Feed state matrix to actor networks and calculate beamforming vectors as explained in Section \ref{sec:method}\;

        Normalize beamforming values to satisfy \eqref{eq:opta} and \eqref{eq:optb}\; 
           
        Observe the reward value \eqref{eq:reward} and store data into $\mathcal{D}$ (in case of using buffer)\;
          \uIf{$t$ $\%$ $\eta$= 0}{
              Sample mini-batch of data from $\mathcal{D}$ or regenerating a new mini-batch of data,
             update actor networks using \eqref{eq:actor loss} or via transfer learning explained in Section \ref{sec:transfer}. \\;
          }
          }
      \uIf{$i$ $\%$ $\eta'$ = 0}{
          Save the model\;}}
  \nonl \textbf{Output}: $\pi_{\boldsymbol{\Omega}}\left(\ba_{b}^{t} \mid \bs_{b}^{t}\right)$ and $\overline{\pi}_{\overline{\boldsymbol{\Omega}}}\left(\ba_{b_0}^{t} \mid \bs_{b_0}^{t}\right)$\;
\caption{{\color{black}Training algorithm for the proposed beamforming method}}
\label{alg:TrainSAC}
\end{algorithm}

\subsection{Testing procedure} \label{sec:test}
The testing phase replicates the training procedure, except that no weight updates are performed. To evaluate the performance of the framework, we load the trained policies $\pi_{\boldsymbol{\Omega}}\left(\ba_{b}^{t} \mid \bs_{b}^{t}\right)$ and $\overline{\pi}_{\overline{\boldsymbol{\Omega}}}\left(\ba_{b_0}^{t} \mid \bs_{b_0}^{t}\right)$ obtained through Algorithm~\ref{alg:TrainSAC}. Following the sequence outlined in Algorithm~\ref{alg:TrainSAC}, each BS applies its corresponding beamforming policy, and the network’s sum rate is recorded at each time slot. The average sum rate performance is presented in Section \ref{sec:results}. This episodic evaluation ensures robustness against variations in user distributions, mobility patterns, and randomness of the sampled channel realizations.

\section{Simulation Results}\label{sec:results}

\subsection{Numerical results}
{\color{black}During training, we configure $B = 4$ clusters, each with a radius of $q = 2$ km and spaced $l = 6$ km apart. Each cluster includes $K = 4$ users uniformly distributed within its area and one LAPS positioned at the center of the cluster at an altitude of 2 km \cite{10379023}, equipped with $N_b = 36$ antennas (for $b \neq b_0$). The HAPS is located at an altitude of 20 km \cite{10445467,shamsabadi2022handling}. To account for realistic placement jitter, at the beginning of each episode, HAPS is randomly positioned within a circle of with a radius of 500 m centered at the network centroid. This captures possible deviations in HAPS placement while preserving coverage symmetry. The HAPS is equipped with $N_{b0} = 64$ antennas.} We then run Algorithm \ref{alg:TrainSAC} for $I_{\rm{episode}} = 200$ episodes, using the Adam optimizer with a learning rate of 0.0004 for both the LAPS and HAPS actor networks.
The system model described in Section~\ref{sec:system model} is implemented in Python, and the proposed entropy-based multi-agent DRL framework is developed using PyTorch. All simulation parameters are summarized in Table~\ref{tab:sim_param}. {\color{black}For the evaluation phase, the results are averaged over 500 test episodes. In this context, we compare the performance of our method against several benchmark schemes, namely: ZF, MRT \cite{10086561}, WMMSE, along with the approach proposed in \cite{ICC}, which replaces the single FNO layer in Fig.~\ref{fig:TBS actor} with two CNN layers. We also include a multi-agent deep deterministic policy gradient (DDPG) baseline for comparison. To achieve its best performance in this setting, the DDPG agent is implemented with a centralized shared critic and a discount factor of 0.3, while our proposed CNN-based and FNO-based actor-only approaches do not require a critic and operate under a fully decentralized execution mode. Importantly, WMMSE, ZF, and MRT are evaluated under perfect CSI and are included only as reference baselines, whereas imperfect CSI is applied exclusively to the learning-based schemes, thus ensuring that classical methods are not disadvantaged by mismatched channel assumptions.}
\begin{table}[h]
\renewcommand{\arraystretch}{1.5}
\caption{{\color{black}Simulation parameters}}
\label{table_example}
\centering
\begin{tabular}{|c||c||c||c|}
\hline
 \textbf{Parameter} & \textbf{Value} & \textbf{Parameter} & \textbf{Value}\\
\hline
 $B$ & 4 & $v$ & 1 m/s \\
\hline
 $N_b$ ($b \neq b_0$) & 36 & $\sigma^{2}$ & -100 dBm\\
\hline
$K$ & 4 & $P_{\max}^{b}$ ($b \neq b_0$)  & 40 watts\\
\hline
$N_{b0}$ & 64 \cite{10445467} & $P_{\max}^{b_0}$ & 100 watts\\
\hline
$\gamma$ & 0.4 & $\gamma'$ & 0.4\\
\hline
$T$ & 50 & $I_{\rm{episode}}$ & 200\\
\hline
$\sigma_{\psi_{d B}}^2$ & 3 dB & learning rate & 0.0004\\
\hline
$c$ & 3 $\times 10^8$ m/s & batch size & 32\\
\hline
$\eta'$ & 10 & $\eta$ & 4\\
\hline
$X$ & 10\cite{alsharoa2020improvement,10379023,10445467} & $T_{c}$ & 0.02 s \\
\hline
$f_{c_{b}}$ ($b \neq b_0$)& 1.8 GHz & $f_{c_{b_0}}$  & 2.7 GHz \\
\hline
\end{tabular}
\label{tab:sim_param}
\end{table}

\begin{figure}[h]
	\centering
	\includegraphics[scale=0.53]{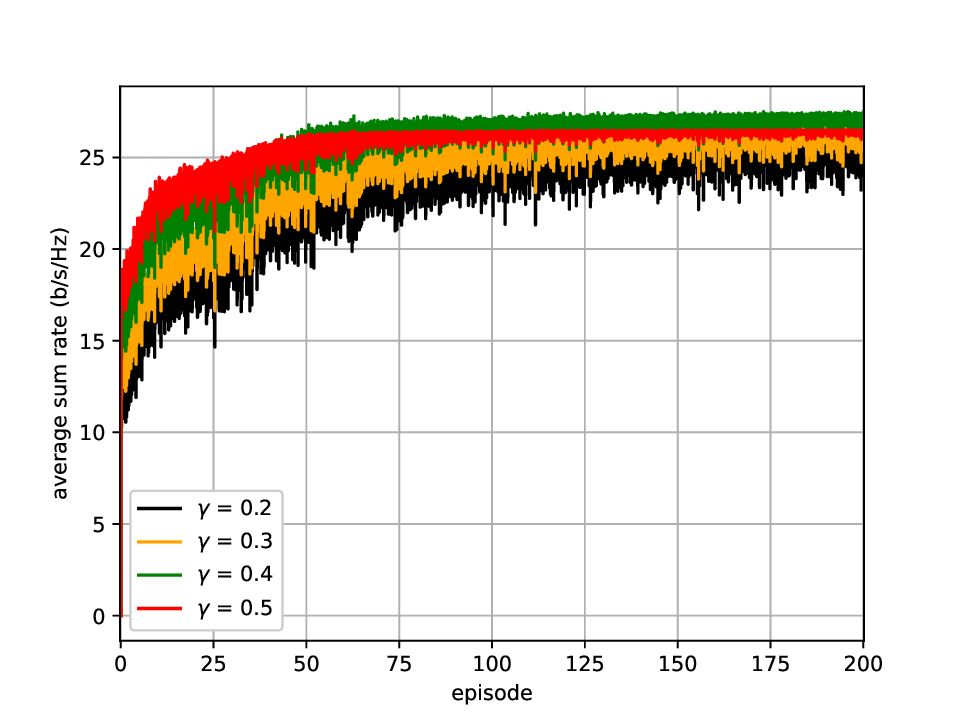}
	\caption{{\color{black}Average sum rate for $K$ = 4 and $B$ = 4 during training for LAPS DNN.}}
	\label{fig:trainhist}
\end{figure}

{\color{black}Fig. \ref{fig:trainhist} illustrates the effect of entropy coefficient $\gamma$ on the average sum rate (for $\xi = 0.6$) during training for the LAPS DNN. It is noted that the reported sum rate represents performance of the whole network (HAPS and LAPS). However, here we show the impact of $\gamma$ for the LAPS actor only (excluding $\gamma'$ for the HAPS actor), since the LAPS DNN handles the more challenging task and, as shown earlier, its learned knowledge can be transferred to the HAPS DNN.
The parameter $\gamma$ controls the level of exploration during training. As can be seen in the results, $\gamma=0.2$ leads to a slower growth in the sum rate and converges to a lower value after 200 episodes. Increasing $\gamma$ enhances exploration and accelerates convergence. However, for $\gamma=0.5$, although the sum rate initially increases faster, it quickly saturates and fails to reach the maximum achievable level. In contrast, $\gamma=0.4$ achieves the best trade-off between exploration and exploitation: specifically, it converges earlier than $\gamma=0.2$ and $\gamma=0.3$ and ultimately attains a higher sum rate compared to the other cases.}

\begin{figure}[h]
	\centering
	\includegraphics[scale=0.53]{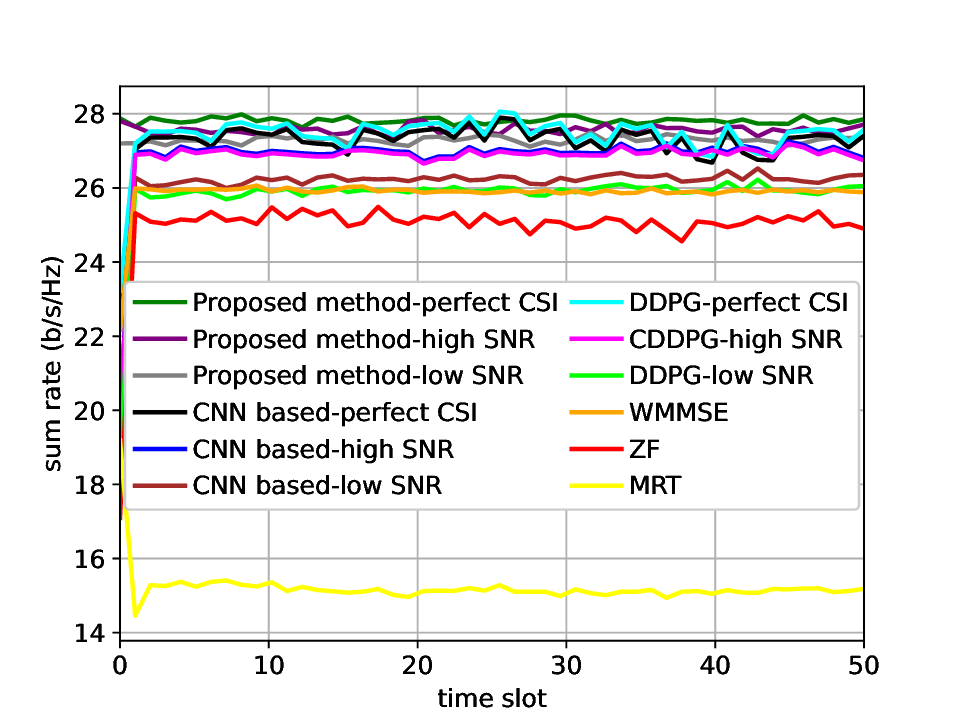}
	\caption{{\color{black}Average sum rate vs. time slot for $K$ = 4 and $B$ = 4.}}
	\label{fig:rate}
\end{figure}

{\color{black}Fig. \ref{fig:rate} shows the average sum rate across time slots. We evaluate our proposed method, the the CNN-based approach from \cite{ICC}, and the DDPG method under the following three CSI conditions: perfect CSI ($\xi = 1$), imperfect CSI at high SNR ($\xi = 0.8$), and imperfect CSI at low SNR ($\xi = 0.6$). These results are also compared to WMMSE, ZF, and MRT techniques evaluated under perfect CSI. For the perfect CSI case, our method achieves average gains of 1.968 bps/Hz, 2.864 bps/Hz, and 12.594 bps/Hz over WMMSE, ZF, and MRT, respectively. Even under the low SNR setting ($\xi = 0.6$), it still outperforms WMMSE and ZF by 1.449 bps/Hz and 2.345 bps/Hz, respectively.
A comparison wth the learning-based baselines reveals that DDPG slightly surpasses the CNN-based method in the perfect CSI case (by about 0.16 bps/Hz). However, under imperfect CSI, the CNN-based method maintains a clear edge, outperforming DDPG by 0.06 bps/Hz at high SNR and 0.3 bps/Hz at low SNR. Notably, at low SNR, WMMSE even outperforms DDPG by 0.056 bps/Hz. By contrast, our FNO-based method remains superior to both CNN and DDPG across all CSI conditions. It is superior to CNN by 0.57 bps/Hz under perfect CSI, by 0.71 bps/Hz at high SNR, and by 1.20 bps/Hz at low SNR; it also outperforms DDPG by 0.41 bps/Hz, 0.77 bps/Hz, and 1.50 bps/Hz under the same scenarios. This difference in performance stems from the fact that, in imperfect CSI, CNNs interpret the noise as local spatial features, leading to a decline in performance as CSI quality deteriorates. Conversely, our FNO-based method filters out high-frequency noise and learns a functional mapping from both perfect and imperfect CSI to beamforming, thus maintaining stable performance even under significant CSI errors. Furthermore, while DDPG outperforms the CNN-based method under perfect CSI at the cost of higher complexity and overhead due to the shared critic, the CNN-based method outperforms DDPG under imperfect CSI due to its entropy-driven exploration and stochastic actors, which makes it more robust to the uncertainty driven by imperfect CSI. The overall performance gain of our proposed method over WMMSE, ZF, and MRT can also be attributed to the enhanced exploration capabilities of the proposed entropy-based multi-agent DRL approach, which effectively performs under various levels of CSI imperfection. Finally, as expected, WMMSE outperforms both ZF and MRT, but does so at the cost of higher computational complexity.}

\begin{figure}[h]
	\centering
	\includegraphics[scale=0.53]{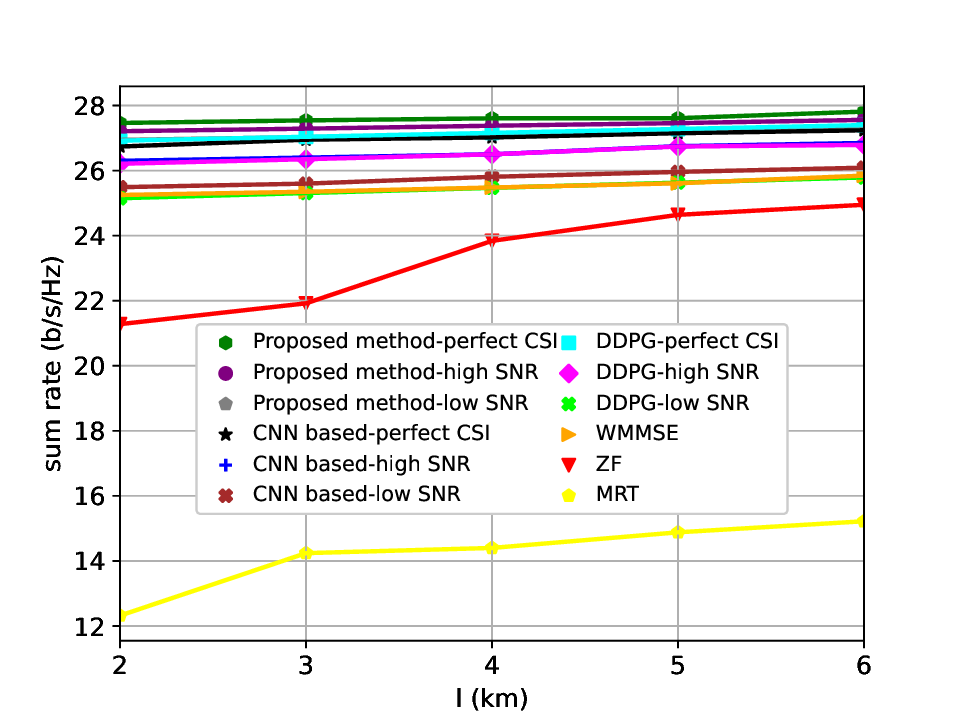}
	\caption{{\color{black}Average sum rate vs. distance between cluster centers for $K$ = 4 and $B$ = 4.}}
	\label{fig:radius}
\end{figure}

{\color{black}Fig.~\ref{fig:radius} illustrates the impact of cluster proximity on the inter-cluster interference and {\color{black}sum rate} performance. As the inter-cluster distance $l$ (i.e., the distance between neighboring cluster centers) approaches the cluster radius $q = 2$ km, users near the cluster boundaries experience a stronger interference from adjacent clusters. Consequently, as clusters become more tightly packed, the network sum rate declines.
Among the evaluated methods, ZF and MRT experience the most significant performance degradation, with corresponding sum rate reductions of 3.67 bps/Hz and 2.9 bps/Hz, respectively, when $l$ decreases from 6 km to 2 km. This is because these schemes are more effective in low-interference regimes. The WMMSE method shows a decrease of 0.596 bps/Hz over the same range. For the CNN-based approach, the sum rate decreases by 0.5 bps/Hz, 0.55 bps/Hz, and 0.6 bps/Hz for the perfect CSI, high SNR ($\xi = 0.8$), and low SNR ($\xi = 0.6$) scenarios, respectively. For the DDPG method, the corresponding degradations are slightly larger: 0.48 bps/Hz, 0.58 bps/Hz, and 0.64 bps/Hz, respectively. In line with Fig. \ref{fig:rate}, while DDPG achieves higher sum rate under perfect CSI, it suffers a greater loss under imperfect CSI, where both the achievable rate and robustness decline compared to the CNN-based method. This behavior stems not only from the absence of entropy-driven exploration, but also from DDPG’s stronger sensitivity to the training environment, which limits its generalization capability. Meanwhile, the proposed FNO-based method exhibits a consistent decline of only 0.35 bps/Hz across all CSI conditions.
Moreover, our method consistently outperforms all benchmarks, even in the low SNR ($\xi = 0.6$) case, which clearly demonstrates its ability to optimally select users and mitigate the effects of increasing inter-cluster interference. This ensures reliable performance across varying values of $l$.
Importantly, both our proposed method and the CNN-based method were trained using $l = 6$ km and evaluated across different values of $l$ without retraining. The generalization capability of our method to unseen inter-cluster distances highlights its robustness. The WMMSE algorithm needs to be re-executed for each new scenario, whereas our method avoids retraining, thereby reducing training complexity.}

\begin{figure}[h]
	\centering
	\includegraphics[scale=0.53]{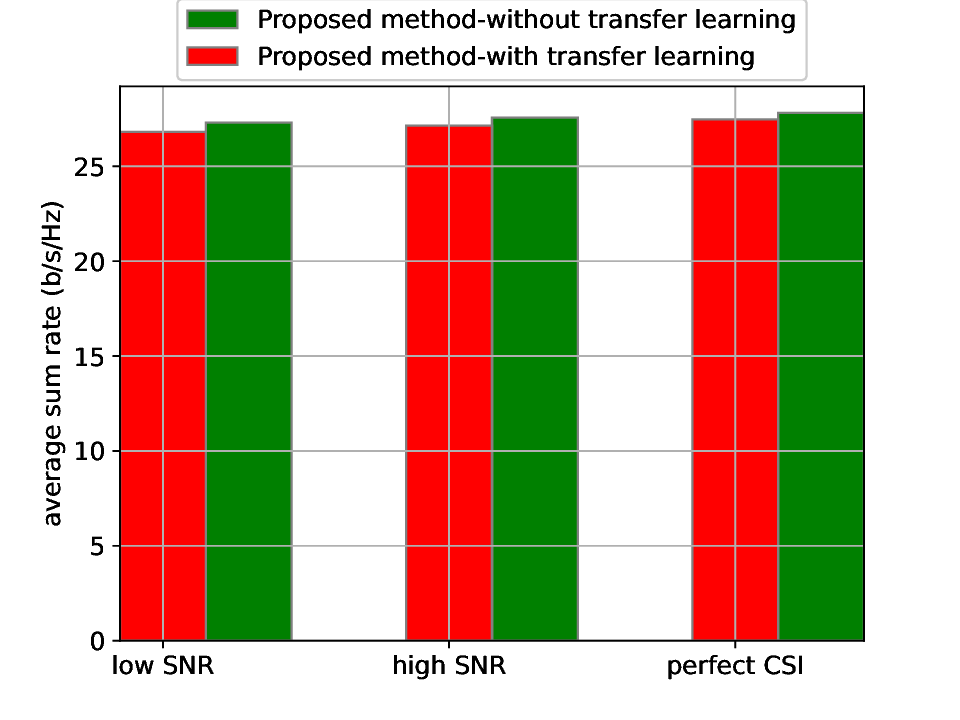}
	\caption{Effect of the proposed transfer learning for $K$ = 4 and $B$ = 4.}
	\label{fig:transfer}
\end{figure}

The proposed system model comprises two aerial layers, where a separate DNN is employed to compute the beamforming vector for each layer's BSs. However, this architecture introduces considerable computational complexity and poses a barrier to scaling multi-layer NTBS deployments in wireless networks. To address this limitation, in Section \ref{sec:transfer}, we propose a transfer learning approach based on a conjugate-prior method. {\color{black}In this strategy, the LAPS’s DNN is treated as the source network, and its learned knowledge is transferred to the HAPS DNN, which serves as the target network, thereby eliminating the need for retraining a separate DNN at each layer.} Specifically, only the FNO layer of the HAPS network is trained, while the hidden and output layers are updated according to \eqref{eq:conjtrf}.
Fig.~\ref{fig:transfer} shows the impact of this transfer learning scheme on the network’s {\color{black}sum rate} performance. As can be seen in Fig.~\ref{fig:transfer}, the proposed method maintains a nearly constant performance under transfer learning, with only minor reductions of 0.35 bps/Hz, 0.42 bps/Hz, and 0.49 bps/Hz for the perfect CSI, high SNR ($\xi = 0.8$), and low SNR ($\xi = 0.6$) scenarios, respectively. Importantly, by avoiding retraining of the hidden and output layers, the computational cost associated with training the HAPS DNN (i.e., the computational complexity of backpropagation) is reduced by nearly 98$\%$ (see Section \ref{sec:com}). This efficiency can be attributed to the strong generalization capability of the FNO and the effective exploration behavior of our entropy-based DRL framework.
Computational complexity of our proposed method, as well as that of \cite{ICC} and WMMSE under various scenarios, is analyzed in further detail in Section \ref{sec:com}.

\begin{figure}[h]
	\centering
	\includegraphics[scale=0.53]{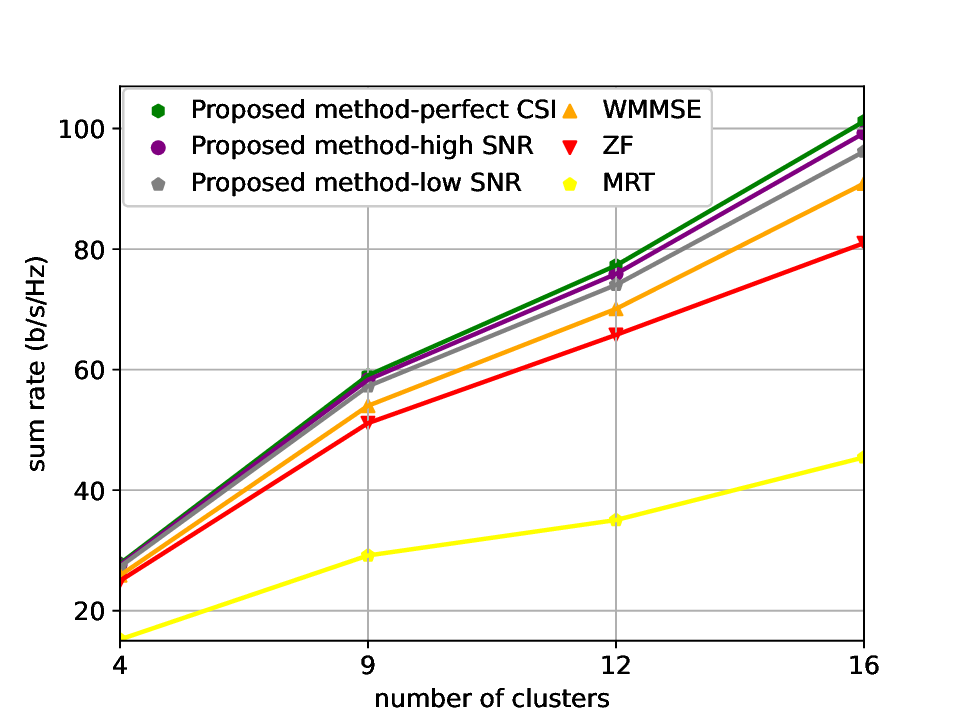}
	\caption{Average sum rate vs. number of clusters for $K$ = 4.}
	\label{fig:cluster}
\end{figure}

NTBSs are designed to serve large geographical areas. To assess the scalability of our proposed method, we evaluate its performance under varying numbers of clusters. Fig.~\ref{fig:cluster} shows the network's sum rate for $B = 4, 9, 12$, and $16$. It is noted that increasing the number of clusters alters the dimensions of both the state and action spaces for $b = b_0$, as defined in \eqref{eq:state} and \eqref{eq:action}, respectively. Consequently, we retrain the HAPS DNN for each cluster configuration. However, since the state and action dimensions for the LAPSs only depend on the number of users per cluster, they remain unchanged. Therefore, the DNN trained for $B = 4$ is reused directly for the cases with $B = 9, 12$, and $16$.
As shown in Fig.~\ref{fig:cluster}, the network's sum rate increases with the number of clusters across all evaluated methods. Between WMMSE and ZF techniques, WMMSE exhibits a greater gain from $B = 12$ to $B = 16$, which could be attributed to its superior handling of the increased inter-cluster interference that impairs ZF’s efficiency. {\color{black}Furthermore, our proposed method, even with a frozen DNN for the LAPSs, demonstrates robust scalability under imperfect CSI conditions.} This confirms its effectiveness in supporting large coverage areas with multiple user clusters. Importantly, reusing the pre-trained LAPS DNN across scenarios eliminates the need for retraining, thereby significantly reducing the computational and time complexities as compared to traditional methods.

\begin{figure}[h]
	\centering
	\includegraphics[scale=0.53]{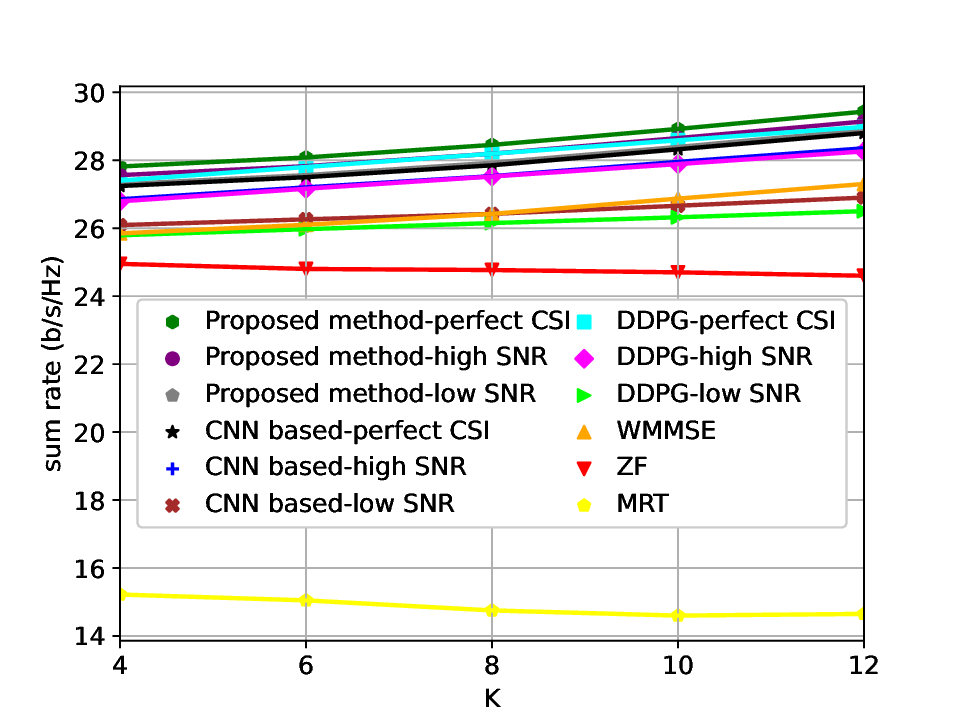}
	\caption{{\color{black}Average sum rate vs. different number of users per cluster for $B$ = 4.}}
	\label{fig:usercluster}
\end{figure}

{\color{black}Fig.~\ref{fig:usercluster} shows the average sum rate as a function of the number of users per cluster. With an increase in $K$ from 4 to 12, the total number of users grows from $U = 16$ to $U = 48$, leading to higher user density and increased intra-cluster interference. This enlarges the action space and complicates effective exploration under imperfect CSI. Consequently, under the low-SNR scenario ($\xi = 0.6$), the CNN-based method struggles to scale with $K$ and beyond $K = 8$ it is outperformed by WMMSE, with a gap of 0.4 bps/Hz at $K = 12$. For this low-SNR case, WMMSE also consistently outperforms DDPG across all values of $K$. Meanwhile, under perfect CSI, DDPG slightly outperforms the CNN-based method, whereas under both high- and low-SNR imperfect CSI, the CNN-based method shows improved performance compared to DDPG. However, in all cases, our proposed method maintains a clear margin over WMMSE, the CNN-based method, and  the DDPG method across all CSI conditions.
Moreover, with an increase in $K$, the performance gap between our method and ZF grows due to the ZF method’s declining efficiency in the high-interference regime (see also Fig.~\ref{fig:radius}). However, our method, exhibits robustness across varying interference conditions and maintains effective exploration as $K$ scales, which explicitly demonstrates its strong scalability with respect to user density.}

\begin{figure}[h]
	\centering
	\includegraphics[scale=0.53]{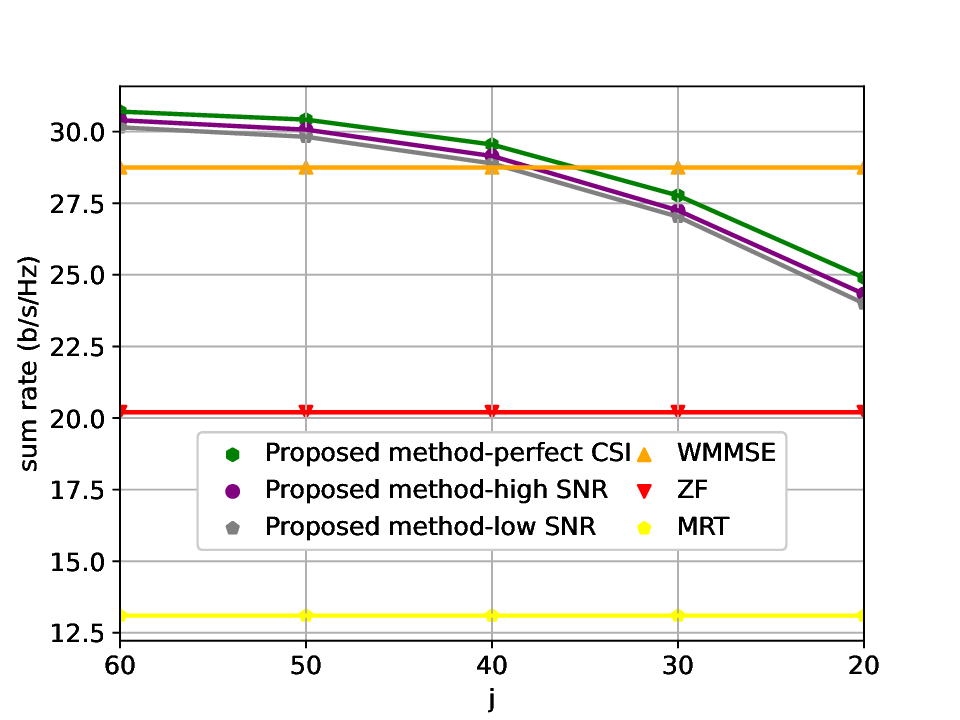}
	\caption{{\color{black}Effect of low rank decomposition on the average {\color{black}sum rate} for $K$ = 20, $B$ = 4 and $N_{b0}$ = 81.}}
	\label{fig:lrd}
\end{figure}

Increasing the number of users in beamforming expands the action space, thus complicating effective action selection and exploration. Concurrently, it enlarges the output dimension of the final layer in Fig. \ref{fig:TBS actor}, necessitating additional hidden layers, which increases the computational complexity, risk of overfitting, and sensitivity to hyperparameter tuning. However, in HAPS-to-ground communications, the strong LoS component induces spatial correlation among user channels, resulting in a beamforming matrix that is not full-rank. Fig. \ref{fig:lrd} examines the effect of the LRD dimension $j$ in \eqref{eqlrd} on the network's sum rate. In this setup, we assume $K = 20$ users per cluster and $N_{b0} = 81$ antennas, while all other parameters are as specified in Table~\ref{tab:sim_param}. WMMSE, ZF, and MRT baselines are evaluated under full-rank beamforming and are included for comparative analysis. To isolate the impact of the LRD, we only apply it on the HAPS level while retaining full-rank beamforming at LAPSs.
The results demonstrate that, when $j = 60$ or $j = 50$, our proposed method maintains a clear performance lead over the WMMSE, indicating that the optimal beamforming matrix remains low-rank under the considered scenario. However, for $j < 50$, the sum rate begins to decline more noticeably. Specifically, comparing $j = 40$ and $j = 30$, the former still outperforms the WMMSE with a less than 1$\%$ reduction in output layer size, whereas the latter reduces the output dimension by 25$\%$ but incurs a 0.97 bps/Hz performance loss relative to WMMSE. Yet, even at $j = 20$, our approach surpasses ZF by 4.7 bps/Hz while reducing output dimensionality by 50 percent. Additional analysis on the computational complexity is provided in Section \ref{sec:com}. Importantly, the optimal choice of $j$ depends on the number of antennas, number of users, and the angular spread of users in the network.

\begin{figure}[h]
	\centering
	\includegraphics[scale=0.53]{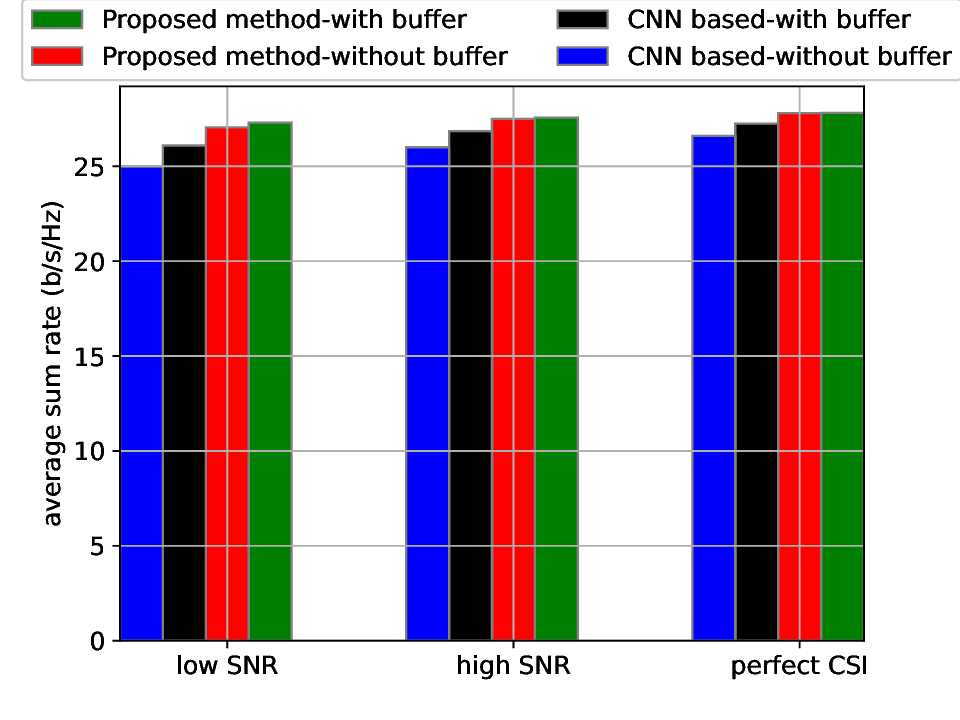}
	\caption{{\color{black}Effect of replay buffer use on average sum rate for $K$ = 4 and $B$ = 4.}}
	\label{fig:buffer}
\end{figure}

As stated in \cite{mnih2015human}, the use of a replay buffer decorrelates data in reinforcement learning and enhances training stability. However, in centralized training with distributed execution—such as in our proposed method—beamforming at each LAPS during the test phase is performed using only its local CSI. This requires sharing the state (i.e., CSI) across agents during the training phase. In Fig. \ref{fig:buffer}, we evaluate the impact of using a replay buffer during training on test-time performance. In scenarios without a buffer, we regenerate channels in \eqref{eq:TBS channel} for randomly located users and use these to form the state matrix in \eqref{eq:state} to update the DNNs. As illustrated in Fig. \ref{fig:buffer}, our proposed method exhibits nearly identical performance under both configurations, offering flexibility for the network operator to balance the trade-off between CSI sharing and artificial data regeneration. In contrast, the CNN-based method experiences performance degradation when trained on regenerated data, as it largely relies on specific numerical patterns in the input. However, the FNO-based approach in our method learns a functional mapping between CSI and beamforming, making it robust to numerical variation and well-suited to training on unseen data. 

To further assess the robustness of our proposed method, we evaluate its performance under the multiplicative imperfect CSI model defined in \eqref{eq:multiplicative} across different user velocities and compare it to the CNN-based method. Both models are trained for $v$ = 1 m/s under the multiplicative CSI model and tested for $v$ = 1, 2, 3, 4, 5, and 6 m/s. Changing $v$ affects the channel values in \eqref{eq:TBS channel} in two key ways. First, higher speeds result in larger variations in $d_{b,u}^{t}$ during an episode, resulting in more pronounced changes in large-scale fading as described in \eqref{eq:largescaleTBS}. This causes the channel to deviate further from the training distribution. Second, increasing $v$ raises the maximum Doppler frequency, which, in turn, decreases the value of $\rho$ in \eqref{eq:HAPS NLOS}, thereby increasing temporal channel variability. Starting from $v$ = 1 m/s, our proposed method experiences a 0.25 bps/Hz drop as compared to the low SNR ($\xi = 0.6$) scenario, whereas the CNN-based method drops by 0.48 bps/Hz. Furthermore, as $v$ increases from 1 m/s to 6 m/s, our method’s performance declines by 0.68 bps/Hz, while the CNN-based method suffers a more substantial degradation of 1.75 bps/Hz. Importantly, this evaluation is conducted without retraining the models for each value of $v$, which represents a challenging test scenario and is aligned with practical generalization requirements.
\begin{figure}[h]
	\centering
	\includegraphics[scale=0.53]{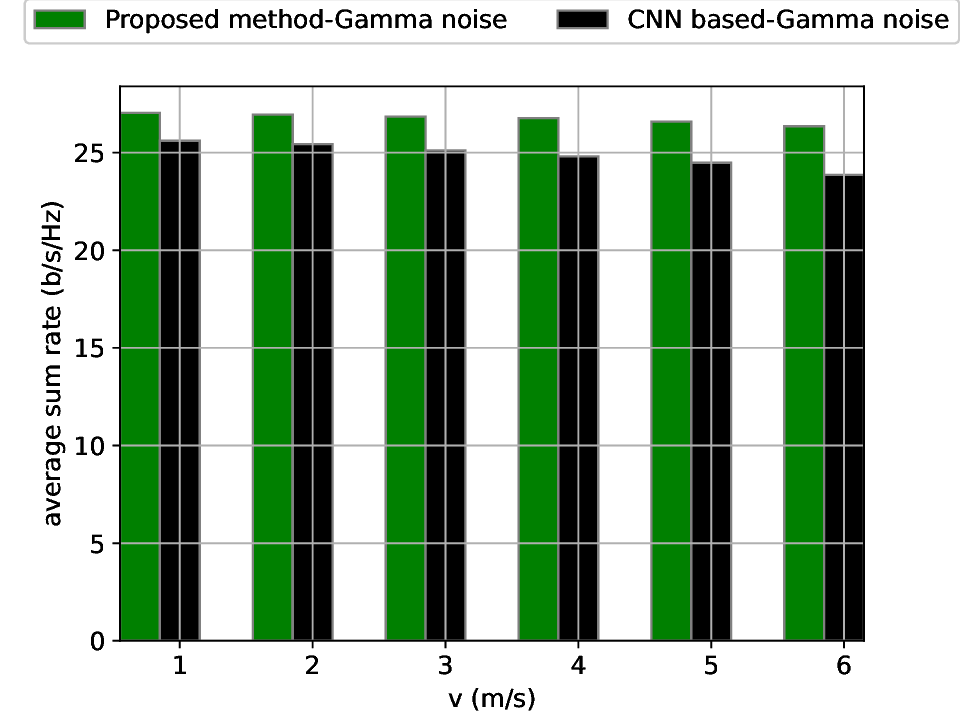}
	\caption{{\color{black}Average sum rate vs. velocity of users $v$ under multiplicative imperfect CSI for $K$ = 4 and $B$ = 4.}}
	\label{fig:velocity}
\end{figure}

{\color{black}\subsection{Computational complexity} \label{sec:com}
The complexity of our proposed method primarily depends on the structure of the DNN. Therefore, in this section, we first compare the complexity of the FNO with that of the CNN-based approach and then evaluate the overall computational burden of our method relative to WMMSE. We also discuss the complexity reduction introduced by our proposed transfer learning and generalization mechanisms. Finally, we evaluate the impact of the LRD on the computational efficiency.

Assuming an input of size $U \times N_{b0}$, i.e., the state dimension for the HAPS, the computational complexity of the FNO layer is $O(2 \times (2 \times U \times N_{b0} \times \text{log}(U) + 2 \times U \times N_{b0} \times \text{log}(N_{b0}) + \overline{L} \times n_{\text{modes}x} \times n_{\text{modes}_y} + \overline{L} \times \overline{O}))$. For a CNN layer with stride 1, the complexity is $O((U - \overline{k} + 1)(N_{b0} - \overline{k} + 1) C_{in} \times C_{out} \times \overline{k}^2)$, where $\overline{k}$ denotes the kernel size, while $C_{in}$ and $C_{out}$ refer to the number of input and output channels, respectively. While the method proposed in \cite{ICC} requires two CNN layers to achieve competitive performance, our approach achieves comparable results using only one FNO layer. For a fully connected layer with input size $l_{in}$ and output size $l_{out}$, the complexity is $O(l_{in} \times l_{out})$.
On the other hand, the WMMSE algorithm involves $\overline{T}$ iterations per time slot, resulting in a complexity of $O(\overline{T}(U^2 \times N_{b0} + U \times N_{b0}^3))$, which rapidly scales with the number of users and antennas.

In our setup, for HAPS beamforming with $U$ = 16 users and $N_{b0}$ = 64 antennas, we use $\overline{T}$ = 100, $n_{\text{modes}x}$ = 20, $n_{\text{modes}_y}$ = 8, $\overline{O}$ = 8, and $\overline{L}$ = 16, resulting in complexities of $O(421068800)$ for WMMSE and $O(5260542)$ for our method. This outcome demonstrates that our method achieves the same task with less than 2$\%$ of the WMMSE complexity. Under the same configuration, the method proposed in \cite{ICC} incurs $O(6808832)$ complexity for the HAPS DNN, meaning that our method achieves approximately 23$\%$ lower computational cost.

Furthermore, our proposed transfer learning framework reduces training complexity of the HAPS DNN by up to 98$\%$, while retaining nearly identical performance (see Fig. \ref{fig:transfer}). In addition, the same DNN trained for LAPSs with $B = 4$ clusters is reused without retraining for $B = 9$, $12$, and $16$, thus effectively rendering the complexity of backward propagation for the LAPSs’ DNN to $O(1)$ for those cases (see Fig. \ref{fig:cluster}).

Finally, as shown in Fig. \ref{fig:lrd}, reducing the rank parameter $j$ affects only the final DNN layers’ dimensions. Although aggressive reduction can degrade performance, choosing a reasonable $j$, such as 60 or 50 (see Fig. \ref{fig:lrd}), can accelerate learning and facilitate exploration in large-scale settings involving many users and antennas.}

{\color{black}\subsection{Communication Overhead Analysis}

In multi-agent DRL frameworks, communication overhead typically arises when agents must exchange state/action information with a centralized critic, or when computing a shared reward requires explicit message passing among agents. Our framework eliminates this critic and relies solely on decentralized actor networks trained with a shared replay buffer. Each APS only observes its own local CSI and generates its beamforming action independently, while the environment computes the shared reward defined in Eq. \eqref{eq:reward} based on the instantaneous rates of all users. This reward is then broadcast to the agents. Consequently, no inter-agent signaling is required to generate the reward. It is noted that the way we analyze the resulting communication overhead follows the methodology in \cite{10934003}.

The only exchange of information occurs during training when transitions $(\bs_{b}^{t}, \ba_{b}^{t}, r^{t} )$ are written into the shared buffer and mini-batches are sampled for policy updates. This operation scales linearly with the number of agents $B$ and batch size $\tilde{D}$, i.e., $O(B\tilde{D})$. Since the replay buffer is maintained in a centralized simulator, as in standard MADRL setups, these operations are local memory accesses, rather than network-level signaling and, therefore, do not contribute to runtime communication overhead during execution. Once training is completed, execution is fully decentralized: each actor computes its beamforming vector using only its own local CSI, without requiring inter-agent communication. By contrast, conventional actor–critic MADRL methods incur an additional $O(B\tilde{D})$ overhead per update step, since each actor must transmit its state–action pairs to a centralized critic and receive gradient information in return. By eliminating the critic, our formulation avoids this extra communication cost, thus effectively halving the training overhead relative to actor–critic approaches.

Furthermore, our framework supports an alternative to the replay-buffer-based training by using synthetic CSI generation. Instead of storing past samples, new channel realizations are regenerated on demand for randomly located users, and the FNO-based actors are updated using these synthetic states. As shown in Fig. \ref{fig:buffer}, this approach achieves nearly identical performance to that of buffer-based training, while CNN-based methods exhibit a noticeable degradation when trained without stored samples. Importantly, synthetic CSI generation eliminates both (i) the need for agents to share CSI during training and (ii) the storage and retrieval overhead of the replay buffer, thereby completely eliminating training communication overhead.

In summary, compared to conventional actor–critic MADRL methods requiring iterative information exchange between actors and a centralized critic, our actor-only formulation significantly reduces communication overhead during training and completely eliminates it during execution. This feature enhances practicality of our framework for large-scale NTN deployments where inter-agent signaling is costly or infeasible.}

\section{Conclusions}
In this study, we proposed a scalable and robust distributed beamforming strategy for a two-layer aerial communication network consisting of HAPS and LAPSs. Using entropy-based multi-agent DRL and FNO, the proposed framework effectively learns a functional mapping from local, imperfect CSI to continuous beamforming vectors. Through the integration of conjugate-prior-based transfer learning and low-rank decomposition, the approach was found to significantly reduce computational complexity while maintaining strong generalization across user mobility, CSI noise, and diverse network topologies. An analysis of communication overhead revealed that the proposed method substantially reduces signaling requirements by (i) eliminating the need for a shared critic during training and (ii) supporting training without a replay buffer via synthetically generated CSI samples. The results of our comprehensive simulations revealed that our method consistently outperforms traditional and AI-based baselines in {\color{black}sum rate} performance, scalability, and robustness, thereby clearly demonstrating its potential for deployment in future multi-layer non-terrestrial networks. For the future work, we envisage investigating scaling to ultra-massive antenna arrays at higher frequencies through modularized state representations \cite{10934003}, incorporating fairness-oriented reward shaping \cite{10934003}, and exploring advanced experience replay strategies \cite{9748970} to further enhance sample efficiency.

\bibliographystyle{IEEEtran}
\bibliography{myreferences}

\begin{IEEEbiography}[{\includegraphics[width=1in,height=1.25in,clip,keepaspectratio]{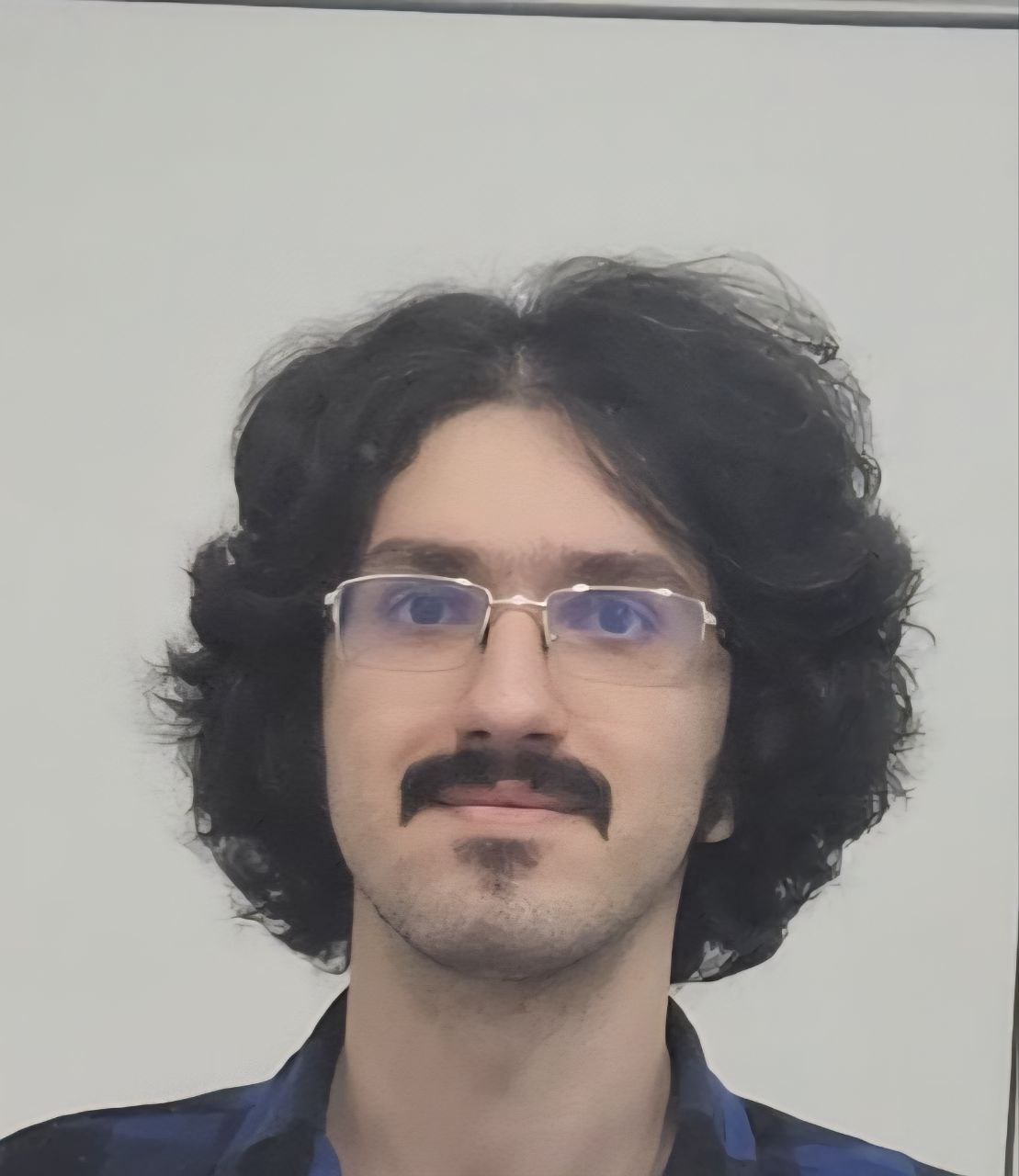}}]{Hesam Khoshkbari} received the B.Sc. degree in electrical engineering from the University of Tabriz, Tabriz, Iran, in 2018, and the M.Sc. degree in electrical engineering from Amirkabir University of Technology (Tehran Polytechnic), Tehran, Iran, in 2021. He is currently pursuing the Ph.D. degree in electrical engineering with École de technologie supérieure (ÉTS), Montreal, QC, Canada. His research interests include AI-enabled wireless communications, HAPS-integrated networks, resource allocation in vertical heterogeneous networks (VHetNets), hybrid beamforming, and deep reinforcement learning for 6G systems.
\end{IEEEbiography}

\begin{IEEEbiography}[{\includegraphics[width=1in,height=1.25in,clip,keepaspectratio]{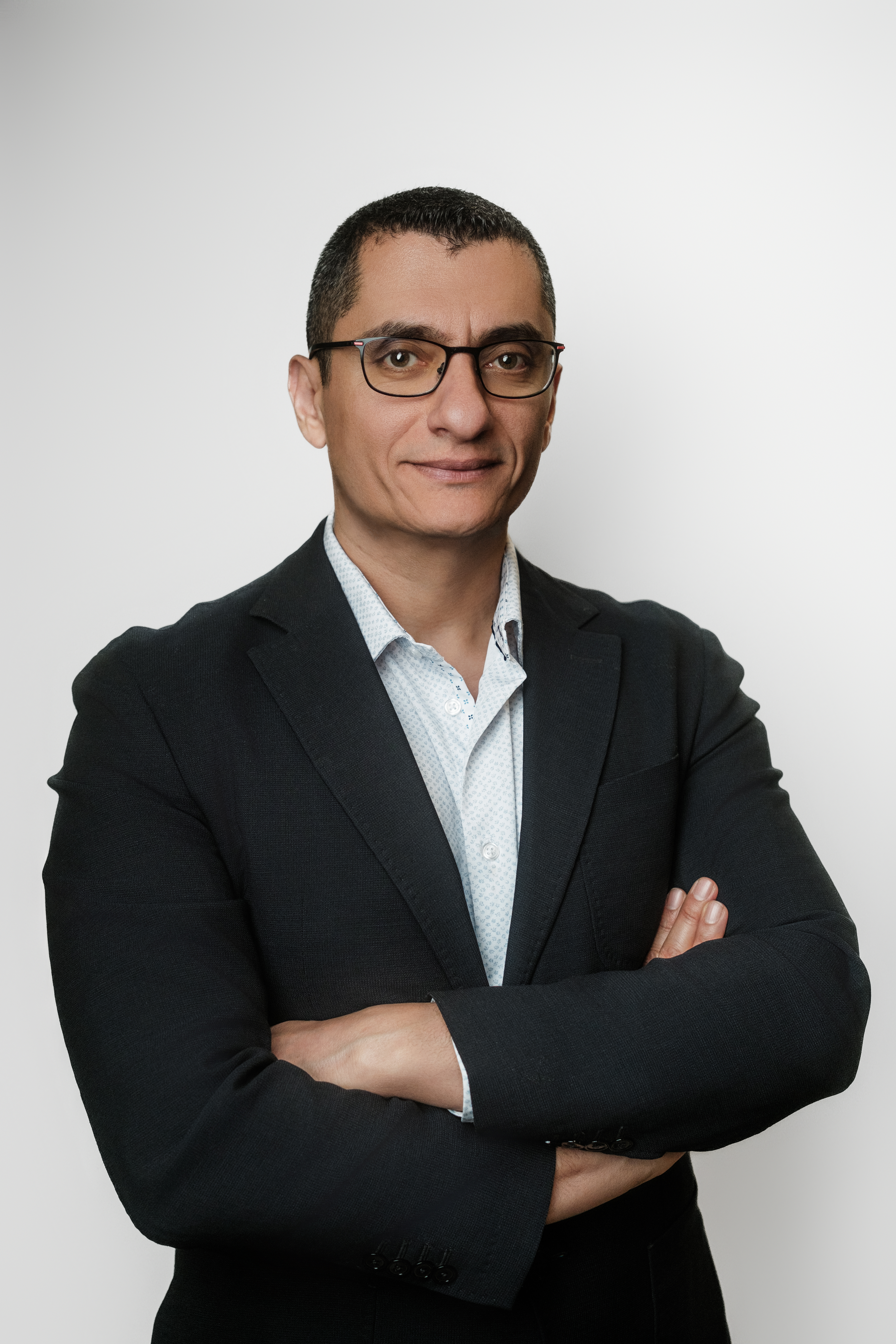}}]{Georges Kaddoum} (Senior Member, IEEE) received the bachelor’s degree in electrical engineering from the École Nationale Supérieure de Techniques Avancées, Brest, France, the M.S. degree in telecommunications and signal processing from the Université de Bretagne Occidentale and Télécom Bretagne, Brest, in 2005, and the Ph.D. (with high honors) degree in signal processing and telecommunications from the National Institute of Applied Sciences, University of Toulouse, Toulouse, France, in 2009. He is currently a Professor and the Research Director of the Resilient Machine Learning Institute, holding the Tier 2 Canada Research Chair with the École de Technologie Supérieure (ÉTS), Canada. He has a prolific publication record with over 300 journal articles and conference papers, two chapters in books, and eight pending patents. His recent research focuses on wireless communication networks, tactical communications, resource allocations, and network security. He has received several accolades, including the Best Paper Awards at prestigious conferences such as IWCMC 2023, PIMRC 2017, and WiMob 2014. Notably, he has been recognized with the IEEE Transactions on Communications Exemplary Reviewer Award in 2019, 2017, and 2015, as well as the Research Excellence Award from the Université du Québec in 2018. In 2019 and 2025, he received the Research Excellence Award from ÉTS. His outstanding contributions were further acknowledged with the 2022 IEEE Technical Committee on Scalable Computing Award for Excellence (Middle Career Researcher), the 2023 MITACS Award for Exceptional Leadership, and the Gold Medal Award from Engineers Canada. He has served as an Associate Editor for the IEEE Transactions on Information Forensics and Security and IEEE Communications Letters. He is currently an Area Editor of the IEEE Transactions on Machine Learning in Communications and Networking, in addition to his role as an Editor for the IEEE Transactions on Communications.
\end{IEEEbiography}

\begin{IEEEbiography}[{\includegraphics[width=1in,height=1.25in,clip,keepaspectratio]{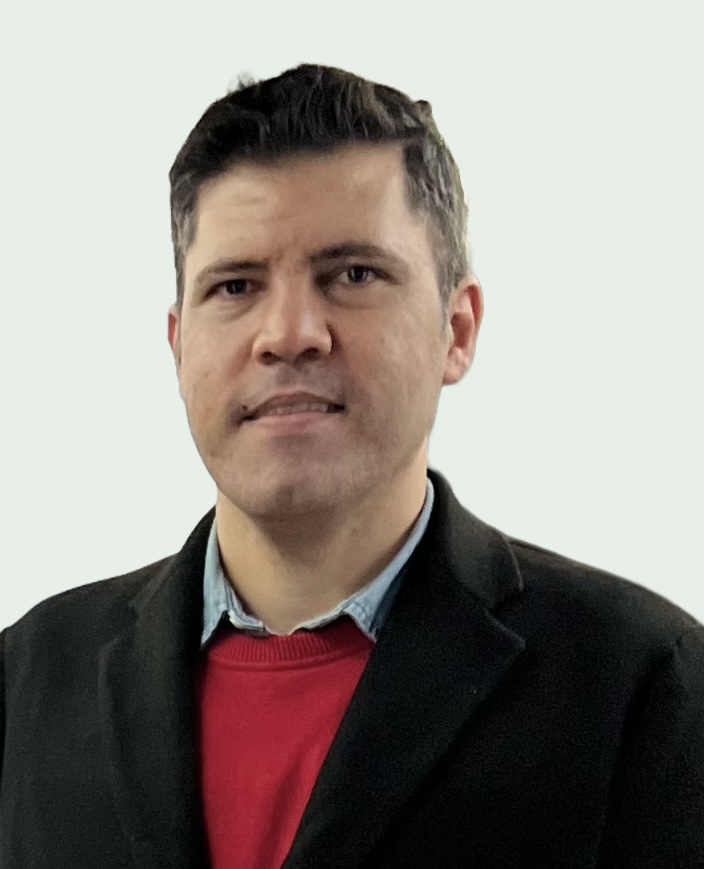}}]{Omid Abbasi} (Senior Member, IEEE) received the B.Sc. degree in electrical engineering from the University of Tabriz, Tabriz, Iran, in 2011, the M.Sc. degree in electrical engineering from the
Amirkabir University of Technology (Tehran Polytechnic), Tehran, Iran, in 2015, and the Ph.D.
degree in electrical engineering from the Sahand University of Technology, Tabriz, in 2020.
From February 2019 to September 2020, he was a Visiting Researcher with the Department of
Systems and Computer Engineering, Carleton University, Ottawa, Canada. Since October 2020,
he has been a Post-Doctoral Fellow with Carleton University. His current research interests
include non-terrestrial networks (NTN) and machine learning for 6G. He has actively served as a
reviewer for flagship IEEE journals and conferences.
\end{IEEEbiography}

\begin{IEEEbiography}[{\includegraphics[width=1in,height=1.25in,clip,keepaspectratio]{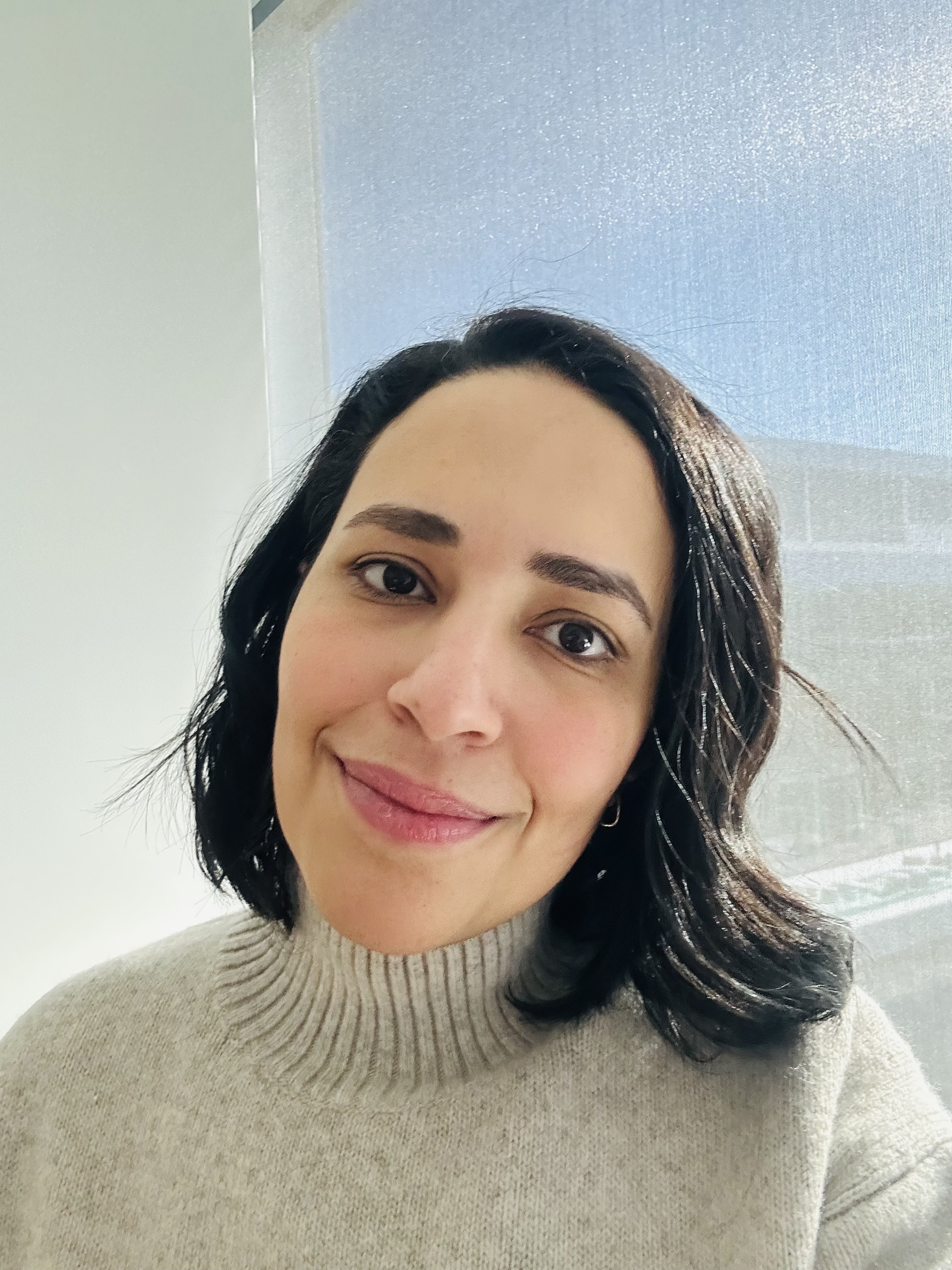}}]{Bassant Selim} received a Master’s degree in communication systems from Pierre et Marie Curie (Paris XI) University, Paris, France, in 2011 and a Ph.D. degree from Khalifa University, Abu Dhabi, United Arab Emirates, in 2017. She is currently Assistant Professor at École de technologie supérieure, Montreal, Canada. Her research interests include intelligent/artificial intelligence-enabled wireless communications, sustainable information and communication technologies, and wireless communications in impulsive environments.
\end{IEEEbiography}

\begin{IEEEbiography}[{\includegraphics[width=1in,height=1.25in,clip,keepaspectratio]{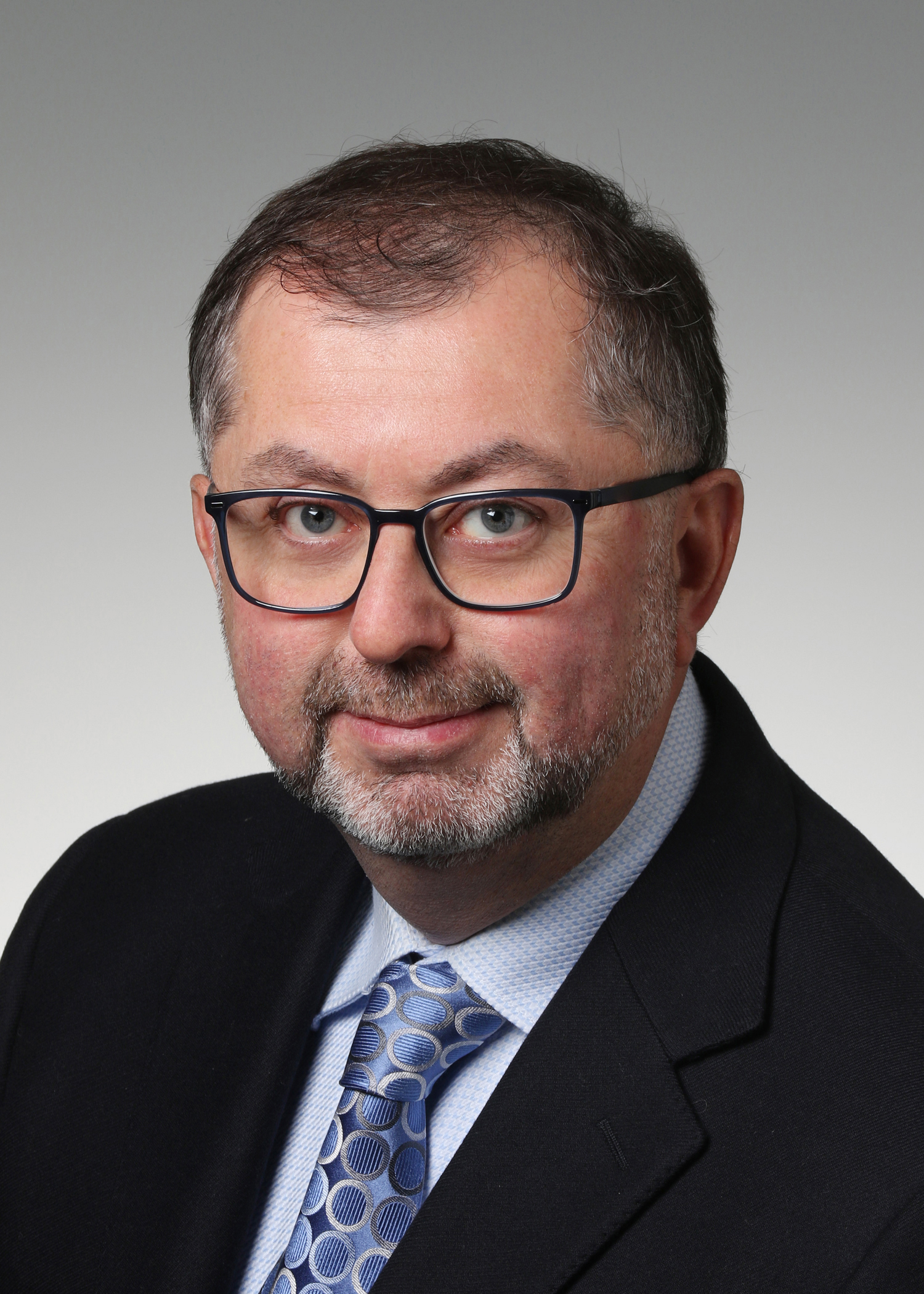}}] {Halim Yanikomeroglu}  is a Chancellor’s Professor in the Department of Systems and Computer
Engineering at Carleton University, Canada; he is also the Founding Director of Carleton-NTN (Non-Terrestrial Networks) Lab. He is a Fellow of IEEE, Engineering Institute of Canada (EIC), Canadian Academy of Engineering (CAE), and Asia-Pacific Artificial Intelligence Association
(AAIA). Dr. Yanikomeroglu has coauthored a high number of papers in 33 different IEEE journals; these papers received tens of thousand of citations. He also has 42 granted patents and impactful technology transfer. He has supervised or hosted at Carleton 180+ postgraduate
researchers; several of his former team members have become professors in Canada, US, UK, and around the world. He gives around 25 invited seminars, keynotes, panel talks, and tutorials every year. He has served as the Steering Committee Chair, General Chair, and Technical
Program Chair of several major international IEEE conferences, as well as in the editorial boards
of several IEEE periodicals. He also served as a Distinguished Speaker for IEEE Communications Society and IEEE Vehicular Technology Society, and an Expert Panelist of the Council of Canadian Academies (CCA|CAC). Dr. Yanikomeroglu received many awards for his research, teaching, and service. He holds a BSc degree in electrical and electronics engineering from the Middle East Technical University (Türkiye), and MASc and PhD degrees in electrical and computer engineering from the University of Toronto (Canada).
\end{IEEEbiography}
\end{document}